    \magnification=1000

   \catcode`\"=12
    \font\black=cmbx10
    \font\sblack=cmbx7 
    \font\ssblack=cmbx5 
    \font\blackital=cmmib10  
    \skewchar\blackital='177 
    \font\sblackital=cmmib7  
    \skewchar\sblackital='177
    \font\ssblackital=cmmib5  
    \skewchar\ssblackital='177 
    \font\sanss=cmss10 
    \font\ssanss=cmss8 scaled 900 
    \font\sssanss=cmss8 scaled 600 
    \font\blackboard=msbm10 
    \font\sblackboard=msbm7 
    \font\ssblackboard=msbm5
    \font\caligr=zplmr7y 
    \font\scaligr=zplmr7y scaled 650 
    \font\sscaligr=zplmr7y scaled 500
     
    \font\fraktur=eufm10 
    \font\sfraktur=eufm7 
    \font\ssfraktur=eufm5

    \font\amsa=msam10

    \font\sya=cmsy10 scaled 650

    \def\all#1{\setbox0=\hbox{\lower1.5pt\hbox{\bsymb
       \char"38}}\setbox1=\hbox{$_{#1}$} \box0\lower2pt\box1\;}
    \def\exi#1{\setbox0=\hbox{\lower1.5pt\hbox{\bsymb \char"39}}
       \setbox1=\hbox{$_{#1}$} \box0\lower2pt\box1\;}

    \def\RP{{\setbox0=\hbox{\sya\char"7B}\box0}}

    \font\bsymb=cmsy10 scaled\magstep2 
    \font\CL=cmcsc10

    \def\tx#1{{\fam0\relax#1}}

    \newfam\bifam 
    \textfont\bifam=\blackital 
    \scriptfont\bifam=\sblackital 
    \scriptscriptfont\bifam=\ssblackital 
    \def\bi#1{{\fam\bifam\relax#1}}

    \newfam\blfam 
    \textfont\blfam=\black 
    \scriptfont\blfam=\sblack 
    \scriptscriptfont\blfam=\ssblack

    \newfam\bbfam 
    \textfont\bbfam=\blackboard 
    \scriptfont\bbfam=\sblackboard 
    \scriptscriptfont\bbfam=\ssblackboard 
    \def\bb#1{{\fam\bbfam\relax#1}}

    \newfam\ssfam 
    \textfont\ssfam=\sanss 
    \scriptfont\ssfam=\ssanss 
    \scriptscriptfont\ssfam=\sssanss 
    \def\ss#1{{\fam\ssfam\relax#1}}

    \newfam\clfam 
    \textfont\clfam=\caligr 
    \scriptfont\clfam=\scaligr 
    \scriptscriptfont\clfam=\sscaligr

    \newfam\frfam 
    \textfont\frfam=\fraktur 
    \scriptfont\frfam=\sfraktur 
    \scriptscriptfont\frfam=\ssfraktur

    \def\hpb#1{\setbox0=\hbox{${#1}$}
    \copy0 \kern-\wd0 \kern.2pt \box0}
    \def\vpb#1{\setbox0=\hbox{${#1}$}
    \copy0 \kern-\wd0 \raise.08pt \box0}
    \def\pmb#1{\setbox0\hbox{${#1}$} \copy0 \kern-\wd0 \kern.2pt \box0} 
    \def\pmbb#1{\setbox0\hbox{${#1}$} \copy0 \kern-\wd0
      \kern.2pt \copy0 \kern-\wd0 \kern.2pt \box0}
    \def\pmbbb#1{\setbox0\hbox{${#1}$} \copy0 \kern-\wd0
      \kern.2pt \copy0 \kern-\wd0 \kern.2pt
    \copy0 \kern-\wd0 \kern.2pt \box0}
    \def\pmxb#1{\setbox0\hbox{${#1}$} \copy0 \kern-\wd0
      \kern.2pt \copy0 \kern-\wd0 \kern.2pt
      \copy0 \kern-\wd0 \kern.2pt \copy0 \kern-\wd0 \kern.2pt \box0}
    \def\pmxbb#1{\setbox0\hbox{${#1}$} \copy0 \kern-\wd0 \kern.2pt
      \copy0 \kern-\wd0 \kern.2pt
      \copy0 \kern-\wd0 \kern.2pt \copy0 \kern-\wd0 \kern.2pt
      \copy0 \kern-\wd0 \kern.2pt \box0}

    \def\blacktriangle{{\setbox0=\hbox{\amsa\char"4E}\box0}} 
    \def\leqslant{{\setbox0=\hbox{\amsa\char"36}\box0}}
    \def\geqslant{{\setbox0=\hbox{\amsa\char"3E}\box0}}

    \mathchardef\za="710B  
    \mathchardef\zb="710C  
    \mathchardef\zg="710D  
    \mathchardef\zd="710E  
    \mathchardef\zve="710F 
    \mathchardef\zz="7110  
    \mathchardef\zh="7111  
    \mathchardef\zvy="7112 
    \mathchardef\zi="7113  
    \mathchardef\zk="7114  
    \mathchardef\zl="7115  
    \mathchardef\zm="7116  
    \mathchardef\zn="7117  
    \mathchardef\zx="7118  
    \mathchardef\zp="7119  
    \mathchardef\zr="711A  
    \mathchardef\zs="711B  
    \mathchardef\zt="711C  
    \mathchardef\zu="711D  
    \mathchardef\zvf="711E 
    \mathchardef\zq="711F  
    \mathchardef\zc="7120  
    \mathchardef\zw="7121  
    \mathchardef\ze="7122  
    \mathchardef\zy="7123  
    \mathchardef\zvp="7124 
    \mathchardef\zvr="7125 
    \mathchardef\zvs="7126 
    \mathchardef\zf="7127  
    \mathchardef\zG="7000  
    \mathchardef\zD="7001  
    \mathchardef\zY="7002  
    \mathchardef\zL="7003  
    \mathchardef\zX="7004  
    \mathchardef\zP="7005  
    \mathchardef\zS="7006  
    \mathchardef\zU="7007  
    \mathchardef\zF="7008  
    \mathchardef\zC="7009  
    \mathchardef\zW="700A  

    \font\kropa=lcircle10 scaled 1700
    \def\tdot{\setbox0=\hbox{\kropa \char"70} \kern1.5pt \raise.35pt \box0}
    \def\sdot{\setbox0=\hbox{\kropa \char"70} \kern1.5pt \raise2.9pt \box0}

    \def\R{{\bb R}}
    

    
    



    \def\sC{{\ss C}}

    \def\sH{{\ss H}}
    \def\sI{{\ss I}}

    \def\sT{{\ss T}}
    
    \def\sV{{\ss V}}

    \def\sr{{\ss r}}
    
    \def\st{{\ss t}}

    \def\rD{{\tx D}}
    \def\rd{{\tx d}}
    \def\tp{{\tilde p}}
    \def\tq{{\tilde q}}
    \def\tf{{\tilde f}}

    \def\wt#1{{\wtilde #1}}
     \def\t#1{{\tilde #1}}

    \def\wP{{\widetilde P}}
    \def\wQ{{\widetilde Q}}
    \def\wS{{\widetilde S}}

    \def\wt{{\widetilde t}}

    \def\oF{{\overline F}}

    \def\oN{{\overline N}}
    \def\oP{{\overline P}}
    \def\oQ{{\overline Q}}
    
    \def\oS{{\overline S}}
    \def\oU{{\overline U}}

    \def\of{{\overline f}}
    \def\oq{{\overline q}}

    \def\o#1{{\overline #1}}

    \def\sPh{{\ss P\ss h}}

    \catcode`\"=\active

    \newcount\secnum  \secnum=0
    \newcount\subsecnum
    \newcount\subsubsecnum

    \newcount\Asecnum  \secnum=0
    \newcount\Asubsecnum

    \newcount\tcount  \tcount=0
    \newcount\pcount  \pcount=0
    \newcount\dcount  \dcount=0
    \newcount\ecount  \ecount=0
    \newcount\ccount  \ccount=0

    \def\Theorem#1{\global\advance\tcount by 1
            \vskip2pt
            \noindent{\CL Theorem \the\tcount.} {\it #1}\par\vskip2pt}

    \def\Proposition#1{\global\advance\pcount by 1
            \vskip2pt
            \noindent{\CL Proposition \the\pcount.} {\it #1}\par\vskip2pt}

    \def\Corollary#1{\global\advance\ccount by 1
            \vskip2pt
            \noindent{\CL Corollary \the\ccount.} {\it #1}\par\vskip2pt}

    \def\Definition#1{\global\advance\dcount by 1
            \vskip2pt
            \noindent{\CL Definition \the\dcount.} #1
            {\hfill \blacktriangle}\par\vskip2pt}

    \def\Example#1{\global\advance\ecount by 1
            \vskip2pt
            \noindent{\CL Example \the\ecount.} #1
            {\hfill \blacktriangle}\par\vskip2pt}

    \def\Proof#1{\noindent{\CL Proof:} #1{\hfill \vrule height 3pt width 5pt depth 2pt}\par\vskip2pt}

    \def\sect#1{
            \global\advance \secnum by 1 \subsecnum=0 \subsubsecnum=0
            \vskip2pt
            \noindent{\bf \the\secnum.\ #1 \vskip-4.5mm}
            \nobreak
            \leftline{}}

    \def\ssect#1{
            \global\advance\subsecnum by 1 \subsubsecnum=0
            \vskip1pt
            \noindent{\bf \the \secnum.\the\subsecnum.\ #1 \vskip-4.5mm}
            \nobreak
            \leftline{}}

    \def\sssect#1{\global\advance\subsubsecnum by 1
            \vskip1pt
            \noindent{\bf \the\secnum.\the\subsecnum.\the\subsubsecnum.\ #1 \vskip-4.5mm}
            \nobreak
            \leftline{}}

    \def\Asect#1{
            \global\advance \Asecnum by 1 \Asubsecnum=0 
            \vskip2pt
            \noindent{\bf \the \Asecnum.\ #1 \vskip-4.5mm}
            \nobreak
            \leftline{}}

    \def\Assect#1{
            \global\advance\Asubsecnum by 1 
            \vskip1pt
            \noindent{\bf A\the \Asecnum .\the\Asubsecnum\  #1 \vskip-4.5mm}
            \nobreak
            \leftline{}}

    \font\Tfont=cmb10 scaled 1300

    \font\tfont=cmb10 

    \def\Title#1{\vskip2mm\centerline{\Tfont #1}\vskip1.5mm}

    \def\title#1{\vskip1mm\leftline{\tfont #1}}

    \def\compose#1#2#3#4#5#6{{\setbox0=\hbox{\raise#2\hbox{\kern#3\hbox{${#1}$}}}
\setbox1=\hbox{\raise#5\hbox{\kern#6\hbox{${#4}$}}}\box0\box1}}

    \def\position#1#2#3{{\setbox0=\hbox{\raise#2\hbox{\kern#3\hbox{${#1}$}}}\box0}}

    \def\Sim#1{\kern.5pt\setbox0=\hbox{$\VPU{8pt}^{\scriptscriptstyle{#1}}$}\setbox1=\hbox{$\sim$}
            \setbox2=\hbox{\kern.5\wd0\copy1\kern-.5\wd0\kern-3pt\copy0}\box2\kern2.7pt}

    \def\RF{\null}
    \def\HEQ#1{{\vbox{\hrule width#1  \vskip1.8pt \hrule width#1}}}
    
    \def\VPU#1{{\vrule height#1 width0pt depth0pt}}
    \def\VPD#1{{\vrule height0pt width0pt depth#1}}

    \def\lpr{{\setbox0=\hbox{\vrule height .15pt width 3.5pt depth 0pt}
\setbox1=\hbox{\vrule height 5.8pt width .3pt depth 0pt}\kern2pt\box0\box1\kern3pt}}

    \def\rpr{{\setbox0=\hbox{\vrule height .15pt width 3.5pt depth 0pt}
\setbox1=\hbox{\vrule height 5.8pt width .3pt depth 0pt}\kern3pt\box0\kern-3.5pt\box1\kern6.5pt}}

    \def\fpr#1{\kern-4pt\setbox0=\hbox{$\VPD{5pt}_{#1}$}\setbox1=\hbox{$\times$}
            \setbox2=\hbox{\kern.5\wd0\copy1\kern-.5\wd0\kern-3pt\copy0}\box2\kern0pt}

    \def\List{\parindent=30pt}
    \def\endList{\parindent=20pt}
    \def\dacapo{\vskip0pt\indent}
    
    \def\Item#1{\item{(#1)} \hskip5pt}

    \def\*{{\textstyle *}}

    \def\polar{{\textstyle\circ}}

 \def\*{{\textstyle *}} \def\s*{{\scriptstyle *}} \def\srP{{\position {\RP}{3pt}{.6pt}}}

    \def\leqs{\kern3pt\leqslant\kern3pt}
    \def\geqs{\geqslant}




   \def\im{{\rm im}}  
    
\def\osT{\sT \position{\scriptscriptstyle\circ}{9pt}{-5.5pt}\kern3pt}

\def\WMT{
        \centerline{W\l odzimierz M. Tulczyjew }
        \centerline{Valle San Benedetto, 2 }
        \centerline{62030 Monte Cavallo, Italy }
        \centerline{Associated with }
        \centerline{Division of Mathematical Methods in Physics}
        \centerline{University of Warsaw}
        \centerline{Ho\.{z}a 74, 00-682 Warszawa}
    \centerline{and}
        \centerline{Istituto Nazionale di Fisica Nucleare,}
        \centerline{Sezione di Napoli}
        \centerline{Complesso Universitario di Monte Sant'Angelo}
        \centerline{Via Cinthia, 80126 Napoli, Italy}
        \centerline{{\tt tulczy@libero.it} }
        \vskip1mm
        }

\def\PUR{
        \centerline{Pawe\l\ Urba\'nski}
        \centerline{Division of Mathematical Methods in Physics}
        \centerline{University of Warsaw}
        \centerline{Ho\.{z}a 74, 00-682 Warszawa}
        \centerline{{\tt urbanski@fuw.edu.pl}}
        \vskip1mm
        }

\catcode`\"=12   
 \catcode`\"=\active

\def\text#1{{\rm\hskip2mm #1 \hskip2mm}} 
 \input pictex
 \input DCpic.sty
 \input xy
 \xyoption{all}
    \hsize=37truepc
    \hoffset=-10truept
    \vsize=52truepc
    \voffset=6truept

    \def\comp{\circ}
     \def\what{\widehat}
     \def\bif{\bi f}


        \Title{Regularity of generating families of functions.}

    \WMT

    \PUR\footnote{}{Research of the second author financed by the Polish Ministry of Science and Higher Education under the grant N N201 365636}
      {\bf Abstract.}
        We describe the geometric structures involved in the variational formulation of physical theories. In presence of these structures,
        the constitutive set of a physical system can be generated by a family of functions. We discuss conditions, under which  a family of functions
        generates an immersed Lagrangian submanifold. These conditions are given in terms of the Hessian of the family.
        \sect{Introduction.}

    The constitutive set of a physical system is frequently a Lagrangian submanifold of a symplectic phase space.  Such
systems are considered {\it reciprocal}.  It is convenient to be able to derive the constitutive set from a simpler {\it generating object} such as a
Lagrangian in the case of dynamics and an internal energy function in the case of statics. The phase space is not usually the cotangent bundle of a
manifold although it is normally isomorphic to a cotangent bundle.  We refer to this isomorphism as a {\it Liouville structure}.  For reasons of
interpretation the Liouville structure can not be used to replace the phase space by the cotangent bundle.  We stress the importance of Liouville
structures for variational formulations of physical theories.  It is the presence of a Liouville structure that permits the generation of a
constitutive set from a generating object.  We say that the system is {\it potential} if its constitutive set is derived from a generatig function or a
function defined on a {\it constraint manifold}. Potentiality implies reciprocity.  A more general generating object, such as a family of functions
does not necesarily generate a Lagrangian submanifold.  We discuss sufficient conditions for families of functions to generate Lagrangian submanifolds.
We define the {\it Hessian} of a family of functions at its critical points. The sufficient conditions for families of functions to generate Lagrangian
submanifolds are based on this definition.

    Reciprocity is an important propery of the constitutive set.  It can be established by examining directly this set.
If the constitutive set is derived from a generating object, then it is more efficient to establish reciprocity by examining the generating object.  A
similar situation arises when conservatiom laws are examined.  Conservation is a property of dynamics and can be established by direct examination of
dynamics.  Noether's theorems simplify the procedure by relating conservation properties to invariance properties of the generating object.

    The paper is organized as follows. In Section 2 we describe some preliminary constructions. In Sections 3 and 4, we describe
geometric structures involved in the variational formulation of physical theories, and the derivation of a set from a generating family of functions.
The notion of a critical point of a family is introduced. Section 5 contains examples of constitutive sets. In Section 6, we recall results concerning
reductions of Lagrangian submanifolds. Then, we discuss the notion of the Hessian of a function (Section 7) and of a family of functions (Section 8),
at a critical point. In Section 9 we introduce the notion of a {\it regular family}, less restrictive then the concept of a Morse family, and we show
that the set generated by a regular family is an immersed Lagrangian submanifold.

        \sect{Preliminary constructions.}
    Let $(P,\zw)$ be a symplectic manifold and let $V$ be a vector subspace of the tangent space $\sT_pP$.
    We denote by $V^\srP$ the {\it symplectic polar}
        $$V^\srP = \left\{\dot p \in \sT_pP ;\; \all{\zd p \in V} \langle \zw, \dot p \wedge \zd p \rangle = 0 \right\}.
                                                                                                             \eqno(1)$$
    If $C \subset P$ is a submanifold, then $\sT^\srP C$ will denote the set
        $$\bigcup_{p \in C} \left(\sT_p C\right)^\srP.
                                                                                                             \eqno(2)$$
    We recall that a submanifold $C \subset P$ is said to be {\it isotropic} if $\sT^\srP C \supset \sT C$.  A
submanifold $C \subset P$ is said to be {\it coisotropic} if $\sT^\srP C \subset \sT C$.  A submanifold $C \subset P$ is said to be {\it Lagrangian} if
$\sT^\srP C = \sT C$.

    A {\it symplectic relation} from a symplectic manifold $(P_1,\zw_1)$ to a symplectic manifold $(P_2,\zw_2)$ is a
differential relation $\zr$ from $P_1$ to $P_2$.  The graph of a symplectic relation is a Lagrangian sybmanifold of the symplectic manifold $(P_2
\times P_1,\zw_2 \ominus \zw_1)$.  The form $\zw_2 \ominus \zw_1$ is defined by
        $$\zw_2 \ominus \zw_1 = pr_2^\* \zw_2 - pr_1^\* \zw_1,
                                                                                                             \eqno(3)$$
    where $pr_1 \,\colon P_2 \times P_1 \rightarrow P_1$ and $pr_2 \,\colon P_2 \times P_1 \rightarrow P_2$ are the
canonical projections.

    If $C$ is a coisotropic submanifold of a symplectic manifold $(P,\zw)$, then the set
        $$D = \left\{\dot p \in \sT P ;\; p = \zt_P(\dot p) \in C, \;\;\all{\zd p \in \sT_p C \subset \sT_p P} \langle
\zw, \dot p \wedge \zd p\rangle = 0 \right\}
                                                                                                             \eqno(4)$$
    is called the {\it characteristic distribution} of the symplectic form $\zw$ restricted to $C$.  At each $p \in C$
the space $D_p = D \cap \sT_p P$ is the symplectic polar $\sT_p^\srP C$ of $\sT_p C$.  The characteristic distribution is Frobenius integrable.  Its
integral manifolds are isotropic submanifolds of $(P,\zw)$ called {\it characteristics} of $\zw|C$.  The set of characteristics may be a manifold
$\wP$.  In this case we introduce the {\it reduction relation} $\zs$ from $P$ to $\wP$.  Its graph is the set
        $${\rm graph }\,(\zs) = \left\{(\tp,p) \in \wP \times P ;\; p \in \tp \right\}.
                                                                                                             \eqno(5)$$
    Let $\zp \,\colon C \rightarrow \wP$ be the canonical projection.  The equality
        $$\zp^\*\wt\zw = \zw|C
                                                                                                             \eqno(6)$$
    defines a symplectic form $\wt\zw$ on $\wP$.  The reduction relation $\zs$ is a symplectic relation from $(P,\zw)$
to $(\wP,\wt\zw)$.  The graph of $\zs$ is the Lagrangian submanifold
        $${\rm graph }\,(\zs) = \left\{(\tp,p) \in \wP \times P ;\; p \in C,\;\; \zp(p) = \tp \right\}.
                                                                                                             \eqno(7)$$
    The projection $\zp$ is the {\it strict symplectic reduction} from $C$ onto the symplectic manifold $(\wP,\wt\zw)$
in the terminology of [1].  It is the essential part of the {\it symplectic reduction relation} $\zs$.

    Let $F$ be a function on a differential manifold $Q$ and let $q \in Q$ be a point.  The differential of $F$ is a
mapping
        $$\rd F \,\colon Q \rightarrow \sT^\*Q.
                                                                                                             \eqno(8)$$
    At $f = \rd F(q) \in \sT^\*Q$ we introduce subspaces
        $$H_f = \sT\rd F(\sT_q Q)
                                                                                                             \eqno(9)$$
        and
        $$V_f = \left\{\zd f \in \sT_f\sT^\*Q ;\; \sT\zp_Q(\zd f) = O_{\zt_Q}(q) \right\}
                                                                                                            \eqno(10)$$
    of the vector space $\sT_f\sT^\*Q$. $O_{\zt_Q}$ is the zero section of $\zt_Q$. The intersection $H_f \cap V_f$ of the subspaces is the subspace
    $\{0\}$ and
the sum $H_f + V_f$ is the entire space $\sT_f\sT^\*Q$.  The subspaces $H_f$ and $V_f$ are images of the injections
        $$i_h \,\colon \sT_q Q \rightarrow \sT_f\sT^\*Q \,\colon \zd q \mapsto \sT\rd F(\zd q)
                                                                                                            \eqno(11)$$
        and
        $$i_v \,\colon \sT_q^\*Q \rightarrow \sT_f\sT^\*Q \,\colon f' \mapsto \st\, Z_{(f,f')}(0),
                                                                                                            \eqno(12)$$
    where $Z_{(f,f')}$ is the curve
        $$Z_{(f,f')} \,\colon \R \rightarrow \sT^\*Q \,\colon s \mapsto f + sf'.
                                                                                                            \eqno(13)$$
    There are also projections
        $$p_h \,\colon \sT_f\sT^\*Q \rightarrow \sT_q Q \,\colon \zd f \mapsto \sT\zp_Q(\zd f)
                                                                                                            \eqno(14)$$
        and
        $$p_v \,\colon \sT_f\sT^\*Q \rightarrow \sT_q^\* Q
                                                                                                            \eqno(15)$$
    such that the mapping
        $$\zC \,\colon \sT_f\sT^\*Q \rightarrow \sT_q Q \oplus \sT_q^\*Q \,\colon \zd f \mapsto p_h(\zd f) \oplus
p_v(\zd f)
                                                                                                            \eqno(16)$$
    is the inverse of
        $$\zF \,\colon \sT_q Q \oplus \sT_q^\*Q \rightarrow \sT_f\sT^\*Q \,\colon \zd q \oplus f' \mapsto i_h(\zd q) +
i_v(f').
                                                                                                            \eqno(17)$$

    The space $\sT_f\sT^\*Q$ is a symplectic vector space with a symplectic form
        $$\zw_f \,\colon \sT_f\sT^\*Q \times \sT_f\sT^\*Q \rightarrow \R
                                                                                                            \eqno(18)$$
    obtained as a restriction of the symplectic form $\zw_Q$ to this vector space.  Both subspaces $H_f$ and $V_f$ are
Lagrangian subspaces.  We choose a pair $(\zd q,f') \in \sT_q Q \times \sT_q^\*Q$ and use curves
        $$\zg \,\colon \R \rightarrow Q
                                                                                                            \eqno(19)$$
        and
        $$\zf \,\colon \R \rightarrow \sT^\*Q
                                                                                                            \eqno(20)$$
    such that $\zd q = \st\zg(0)$, $\zf(0) = f'$, and $\zp_Q \comp \zf = \zg$.  The mapping
        $$\zq \,\colon \R^2 \rightarrow \sT^\*Q \,\colon (s_1,s_2) \mapsto Z_{((\rd F \comp \zg)(s_1),\zf(s_1))}(s_2)
                                                                                                            \eqno(21)$$
    represents the pair
        $$(i_h(\zd q),i_v(f')) \in \sT_f\sT^\*Q \times \sT_f\sT^\*Q
                                                                                                            \eqno(22)$$
    in the sense that
        $$\st\zq(\cdot,0)(0) = i_h(\zd q)
                                                                                                            \eqno(23)$$
        and
        $$\st\zq(0,\cdot)(0) = i_v(f').
                                                                                                            \eqno(24)$$

    In the following calculation we use the facts that $\zw_Q$ is the differential of the Liouville form $\zy_Q$, that
the Liouville form is vertical and that for each $s$ the curve $\zq(s,\cdot) \,\colon \R \rightarrow \sT^\*Q$ is vertical.
        $$\eqalign{
    \zw_f(i_h(\zd q),i_v(f')) &= \langle \zw_Q, i_h(\zd q) \wedge i_v(f') \rangle \cr
    &= {{\rd}\over{\rd s}}\langle \zy_Q, \st\zq(s,\cdot)(0)\rangle\big|_{s=0} - {{\rd}\over{\rd s}}\langle \zy_Q,
\st\zq(\cdot,s)(0)\rangle\big|_{s=0} \cr
    &= - {{\rd}\over{\rd s}}\langle \zy_Q, \st\zq(\cdot,s)(0)\rangle\big|_{s=0} \cr
    &= - {{\rd}\over{\rd s}}\langle \zq(0,s), \sT\zp_Q(\st\zq(\cdot,s)(0))\rangle\big|_{s=0} \cr
    &= - {{\rd}\over{\rd s}}\langle f + sf', \st(\zp_Q \comp \zq(\cdot,s))(0)\rangle\big|_{s=0} \cr
    &= - {{\rd}\over{\rd s}}\langle f + sf', \zd q\rangle\big|_{s=0} \cr
    &= - \langle f', \zd q\rangle.  }
                                                                                                            \eqno(25)$$

    The formula
        $$\eqalign{
    \zw_f(i_h(\zd_1 q) + i_v(f_1'),i_h(\zd_2 q) + i_v(f_2')) &= \zw_f(i_h(\zd_1 q),i_v(f_2')) +
\zw_f(i_v(f_1'),i_h(\zd_2 q)) \cr
    &= \langle f_1', \zd_2 q\rangle - \langle f_2', \zd_1 q\rangle \cr
    }
                                                                                                            \eqno(26)$$
    shows that the mapping \RF(17) is a linear symplectomorphism from the direct product $\sT_q Q \oplus \sT_q^\*Q$ with
its canonical symplectic structure to the symplectic vector space $(\sT_f\sT^\*Q,\zw_f)$.  The formula
        $$\zw_f(\zd_1 f,\zd_2 f) = \langle p_v(\zd_1 f), p_h(\zd_2 f)\rangle - \langle p_v(\zd_2 f), p_h(\zd_1 f)\rangle
                                                                                                            \eqno(27)$$
    is equivalent to \RF(26).

        \sect{Subsets of symplectic manifolds generated by families.}
    The geometric structures involved in the variational formulation of a physical theory are represented by the diagram

    \vskip1mm
        $$\vcenter{
    \begindc{0}[1]
    \obj(000,055)[01]{$(P,\zw)$}
    \obj(080,055)[11]{$(\sT^\*Q,\zw_Q)$}
    \obj(150,055)[21]{$\oQ$}
    \obj(210,000)[30]{$$}
    \obj(210,055)[31]{$\R$}
    \obj(000,000)[00]{$Q$}
    \obj(080,000)[10]{$Q$}
    \obj(150,000)[20]{$\wQ$}
    \mor{01}{00}[10,10]{$\zp$}[0,0]
    \mor{01}{11}[10,10]{$\za$}[1,0]
    \mor{20}{10}[10,10]{$\zi$}[-1,0]
    \mor{11}{10}[10,10]{$\zp_Q$}[0,0]
    \mor{21}{31}[10,10]{$\oU$}[1,0]
    \mor{21}{20}[10,10]{$\zh$}[0,0]
    \obj(040,000)[99|00|10]{$\HEQ{052pt}$}
    \enddc}
                                                                                                            \eqno(28)$$
    \vskip2mm

    The object $(P,\zw)$ is the {\it phase space} of the theory.  The diagram
    \vskip1mm
        $$\vcenter{
    \begindc{0}[1]
    \obj(000,055)[01]{$(P,\zw)$}
    \obj(000,000)[00]{$Q$}
    \mor{01}{00}[10,10]{$\zp$}[0,0]
    \enddc}
                                                                                                            \eqno(29)$$
    \vskip2mm
    \noindent is a vector fibration projecting the phase space onto the {\it configuration space} $Q$ and the diagram

    \vskip1mm
        $$\vcenter{
    \begindc{0}[1]
    \obj(000,055)[01]{$(P,\zw)$}
    \obj(080,055)[11]{$(\sT^\*Q,\zw_Q)$}
    \obj(000,000)[00]{$Q$}
    \obj(080,000)[10]{$Q$}
    \obj(40,000)[30]{$\HEQ{22mm}$}
    \mor{01}{00}[10,10]{$\zp$}[0,0]
    \mor{01}{11}[10,10]{$\za$}[1,0]
    \mor{11}{10}[10,10]{$\zp_Q$}[0,0]
    \enddc}
                                                                                                            \eqno(30)$$
    \vskip2mm
    \noindent is a vector fibration isomorphism establishing a {\it Liouville structure} for the phase space $(P,\zw)$.
The remaining part
    \vskip1mm
        $$\vcenter{
    \begindc{0}[1]
    \obj(000,055)[01]{$$}
    \obj(060,055)[11]{$\oQ$}
    \obj(120,055)[21]{$\R$}
    \obj(000,000)[00]{$Q$}
    \obj(060,000)[10]{$\wQ$}
    \obj(120,000)[20]{$$}
    \mor{10}{00}[10,10]{$\zi$}[-1,0]
    \mor{11}{10}[10,10]{$\zh$}[-1,0]
    \mor{11}{21}[10,10]{$\oU$}[1,0]
    \enddc}
                                                                                                            \eqno(31)$$
    \vskip2mm
    \noindent is a {\it generating object}.  It consists of the injection

    \vskip1mm
        $$\vcenter{
    \begindc{0}[1]
    \obj(000,000)[00]{$Q$}
    \obj(060,000)[10]{$\wQ$}
    \mor{10}{00}[10,10]{$\zi$}[-1,0]
    \enddc}
                                                                                                            \eqno(32)$$
    \vskip2mm
    \noindent of a submanifold $\wQ \subset Q$, a differential fibration

    \vskip1mm
        $$\vcenter{
    \begindc{0}[1]
    \obj(000,055)[01]{$\oQ$}
    \obj(000,000)[00]{$\wQ$}
    \mor{01}{00}[10,10]{$\zh$}[-1,0]
    \enddc}
                                                                                                            \eqno(33)$$
    \vskip2mm
    \noindent and a function $\oU \,\colon \oQ \rightarrow \R$ interpreted as a family of functions defined on fibres of
the fibration $\zh$ and denoted by $(\oU,\zh)$.  The generating object generates a subspace of the phase space.  There is an alternate representation
of the Liouville structure in terms of a pairing
        $$\langle \;\, ,\;\rangle \,\colon P \fpr{(\zp,\zt_Q)} \sT Q \rightarrow \R.
                                                                                                            \eqno(34)$$
    defined by
        $$\langle \za(p), v\rangle\VPD{4pt}_Q = \langle p, v\rangle
                                                                                                            \eqno(35)$$
    for $p \in P$ and each $v \in \sT Q$ such that $\zt_Q(v) = \zp(p)$.  The canonical pairing

        $$\langle \;\, ,\;\rangle\VPD{4pt}_Q \,\colon \sT^\*Q \fpr{(\zp_Q,\zt_Q)} \sT Q \rightarrow \R
                                                                                                            \eqno(36)$$
    is used.  The relation \RF(35) defines the pairing \RF(34) in terms of the symplectomorphism $\za$ or the
symplectomorphism in terms of the pairing.  The set

        $$\eqalign{
    S &= \left\{p \in P;\; \tq = \zp(p) \in \wQ, \exi{\oq \in \zh^{-1}(\tq)} \all{\zd\oq \in \sT_\oq \oQ}\right.  \cr
    & \hskip30mm \left.  \zd\tq = \sT\zh(\zd\oq) \Rightarrow \langle p, \zd\tq \rangle = \langle \rd \oU,
\zd\oq \rangle \right\}
    }
                                                                                                            \eqno(37)$$
    generated by the generating object \RF(31) is expected to be a Lagrangian submanifold of the phase space $(P,\zw)$.

    The formula \RF(37) defines the set $S$ directly in terms of the generating object.  There is an alternate
derivation of this set by the following sequence of operations.
        \List
    \Item{1} The function $\oU$ is used to generate the Lagrangian submanifold $\oS = \im(\rd\oU)\subset \sT^\*\oQ$ of
the symplectic manifold $(\sT^\*\oQ,\zw_\oQ)$.
    \Item{2} The {\it phase lift} symplectic relation
        $$\sPh\,\zh \,\colon (\sT^\*\oQ,\zw_\oQ) \rightarrow (\sT^\*\wQ,\zw_\wQ)
                                                                                                            \eqno(38)$$
    of the fibration $\zh$ is used to produce the set $\wS = \sPh\,\zh(\oS) \subset \sT^\*\wQ$. The relation $\sPh\,\zh$
    can be described in the following way. We denote by $\sV\oQ$ the subbundle
        $$\left\{\zd\oq \in \sT\oQ ;\; \sT\zh(\zd\oq) = 0 \right\}
                                                                                                            \eqno(39)$$
    of the tangent bundle $\sT\oQ$ composed of {\it vertical} vectors.  The {\it polar}
        $$\sV^\polar \oQ = \left\{\of \in \sT^\*\oQ;\; \all{\zd\oq \in \sV\oQ} \zt_\oQ(\zd\oq) = \zp_\oQ(\of)
\Rightarrow \langle \of, \zd\oq\rangle = 0 \right\}
                                                                                                            \eqno(40)$$
    of this {\it vertical subbundle} is a coisotropic submanifold of $(\sT^\*\oQ,\zw_\oQ)$.  Let $\of \in
\sV^\polar\oQ$, $\oq = \zp_\oQ(\of)$, and $q = \zh(\oq)$.  The relation
        $$\langle \t\zh(\of), \zd q\rangle = \langle \of, \zd\oq\rangle
                                                                                                            \eqno(41)$$
    with $\zd q \in \sT_q Q$ and $\zd\oq \in \sT_\oq\oQ$ such that $\sT\zh(\zd\oq) = \zd q$

    defines a differential fibration
        $$\t\zh \,\colon \sV^\polar \oQ \rightarrow \sT^\*\wQ.
                                                                                                            \eqno(42)$$
    \item{}
    This fibration is the strict symplectic reduction (see [1]) from $\sV^\polar\oQ$ onto the symplectic manifold
$(\sT^\*\wQ,\zw_\wQ)$.  It is the essential part of the symplectic reduction relation \RF(38) whose graph is the set
        $$\left\{(\tf,\of) \in \sT^\*\wQ \times \sT^\*\oQ;\; \of \in \sV^\polar \oQ, \tf = \t\zh(\of) \right\}.
                                                                                                            \eqno(43)$$
    The reduced set
        $$\wS = \sPh\,\zh(\oS) = \t\zh(\oS \cap \sV^\polar \oQ)
                                                                                                            \eqno(44)$$
    is not necessarily a Lagrangian submanifold.
    \Item{3} The phase lift
        $$\sPh\,\zi \,\colon (\sT^\*\wQ,\zw_\wQ) \rightarrow (\sT^\*Q,\zw_Q)
                                                                                                            \eqno(45)$$
    of the injection $\zi$ is applied to the set $\wS$.  The result is the set $\what S = \sPh\,\zi(\wS) \subset
\sT^\*Q$. The relation $\sPh\,\zi$ is, essentially, the strict symplectic reduction from a coisotropic submanifold $\zp_Q^{-1}(\zi(\wQ))$ of
$(\sT^\*Q,\zw_Q)$ onto $(\sT^\*\wQ,\zw_\wQ)$. This reduction is the mapping
        $$\hat\zi \,\colon \zp_Q^{-1}(\zi(\wQ)) \rightarrow \sT^\*\wQ
                                                                                                            \eqno(46)$$
    characterized by
        $$\langle \hat\zi(f), \zd \tq\rangle = \langle f, \sT\zi(\zd \tq)\rangle
                                                                                                            \eqno(47)$$
    for each $\zd \tq \in \sT_{\tq} \wQ$, \ ${\zi(\tq)=\zp_Q(f)}$.  If $\wS$ is a Lagrangian submanifold of
$(\sT^\*\wQ,\zw_\wQ)$, then
        $$\what S = \sPh\,\zi(\wS) = \hat\zi^{-1}(\wS \cap \zp_Q^{-1}(\wQ))
                                                                                                            \eqno(48)$$
    is a Lagrangian submanifold of $(\sT^\*Q,\zw_Q)$,
    \Item{4} The set $S \subset \sT^\*Q$ is finally obtained as the inverse image $\za^{-1}(\what S)$.  This set is a
Lagrangian submanifold of $(P,\zw)$ if $\what S$ is a Lagrangian submanifold of $(\sT^\*Q,\zw_Q)$.
        \endList
    \vskip2mm

    In the following example we have a nontrivial Liouville structure and a constrained generating family although the
constraint is open.

        \Example{
    Let $M$ be the space time of general relativity with a Minkowski metric $g \,\colon \sT M \rightarrow \sT^\*M$ of
signature $(1,3)$.  The Lagrangian of a free particle of mass $m$ is the function
        $$L \,\colon \wQ \rightarrow \R \,\colon \dot x \mapsto m\sqrt{\langle g(\dot x), \dot x\rangle}
                                                                                                            \eqno(49)$$
    defined on the open submanifold
        $$\wQ = \left\{\dot x \in \sT M ;\; \langle g(\dot x), \dot x \rangle > 0 \right\}
                                                                                                         $$
    of time-like vectors in $Q = \sT M$.  The dynamics of the particle is a differential equation in the energy-momentum
phase space $\sT^\*M$.  It is therefore a subset $D \subset \sT\sT^\*M$.  The space $\sT\sT^\*M$ has a natural symplectic structure.
    The symplectic form is the total differential $\rd_T\zw_M$ of the canonical symplectic form $\zw_M$ in $\sT^\*M$.
    $\rd_T$ is a derivation  on the exterior algebra of forms on a manifold $M$ with
values in the exterior algebra of forms of the tangent bundle $\sT M$ (for definition see, e.g.,[2]). The dynamics is a Lagrangian submanifold of
$(\sT\sT^\*M,\rd_T\zw_M)$.  The Liouville structure
    \vskip1mm
        $$\vcenter{
    \begindc{0}[1]
    \obj(000,060)[01]{$\left(\sT\sT^\*M,\rd_T\zw_M\right)$}
    \obj(125,060)[11]{$\left(\sT^\*\sT M,\zw_{\sT M}\right)$}
    \obj(000,000)[00]{$\sT M$}
    \obj(125,000)[10]{$\sT M$}
    \obj(062,000)[99|00|30]{$\HEQ{100pt}$}
    \mor{01}{00}[10,10]{$\sT\zp_M$}[0,0]
    \mor{01}{11}[10,10]{$\za_M$}[1,0]
    \mor{11}{10}[10,10]{$\zp_{\sT M}$}[0,0]
    \enddc}
                                                                                                            \eqno(51)$$
    \vskip2mm
    \noindent is used for generating dynamics from the Lagrangian \RF(49).  This Liouville structure was introduced in
[3].  It is described rigorously in [4].  At each phase $p \in \sT^\*M$ the pseudoriemannian structure of $M$ defines subspaces $H_p \subset
\sT_p\sT^\*M$ and $V_p \subset \sT_p\sT^\*M$ of horizontal and vertical vectors such that
        $$\sT_p\sT^\*M = H_p + V_p
                                                                                                            \eqno(52)$$
        $$H_p \cap V_p = \{O_{\zt_{\sT^*M}}(p)\}.
                                                                                                            \eqno(53)$$
    The dynamics is the set
        $$D = \left\{\dot p \in \sT\sT^\*M;\; \sT\zp_M(\dot p) \in \wQ, \;\;\zt_{\sT^*M}(\dot p) = {{mg(\sT\zp_M(\dot
p))}\over{\|\sT\zp_M(\dot p)\|}},\;\;\dot p \in H_{\zt_{\sT^*M}(\dot p)} \right\}
                                                                                                            \eqno(54)$$
    with
        $$\|\sT\zp_M(\dot p)\| = \sqrt{\langle g(\sT\zp_M(\dot p)), \sT\zp_M(\dot p)\rangle }
                                                                                                            \eqno(55)$$
        }

        \Example{
    The Hamiltonian generating object for the dynamics of Example~1 
        $$H \,\colon \oP \rightarrow \R \,\colon (p,\zl) \mapsto \zl(\sqrt{\langle p, g^{-1}(p)\rangle} - m)
                                                                                                            \eqno(56)$$
    is defined on $\oP = \wP \times \R_+$, where $\wP$ is the set
        $$\wP = \left\{p \in \sT^\*M ;\; \langle p, g^{-1}(p)\rangle > 0 \right\}
                                                                                                         $$
    is treated as a family of functions on fibres of the projection
        $$\zz \,\colon \oP \rightarrow \wP \,\colon (p,\zl) \mapsto p.
                                                                                                            \eqno(58)$$
    The Liouville structure
    \vskip1mm
        $$\vcenter{
    \begindc{0}[1]
    \obj(000,060)[01]{$\left(\sT\sT^\*M,\rd_T\zw_M\right)$}
    \obj(140,060)[11]{$\left(\sT^\*\sT^\*M,\zw_{\sT^\*M}\right)$}
    \obj(000,000)[00]{$\sT^\*M$}
    \obj(140,000)[10]{$\sT^\*M$}
    \obj(070,000)[99|00|30]{$\HEQ{110pt}$}
    \mor{01}{00}[10,10]{$\zt_{\sT^\*M}$}[0,0]
    \mor{01}{11}[10,10]{$\zb_{(\sT^\*M,\zw_M)}$}[1,0]
    \mor{11}{10}[10,10]{$\zp_{\sT^\*M}$}[0,0]
    \enddc}
                                                                                                            \eqno(59)$$
    \vskip2mm
    \noindent is used.
        }

        \Example{
    Let $M$ be the space-time manifold of General Relativity.  It is a pseudoriemannian manifold of dimenion 4 with a
metric tensor $g \,\colon \sT M \rightarrow \sT^\*M$.
    \dacapo
    The Lagrangian generating family for the dynamics of a massless particle is the function
        $$L \,\colon \oQ \rightarrow \R \,\colon (\dot x,\zm) \mapsto {{1}\over{2\zm}} \langle g(\dot x), \dot x\rangle
                                                                                                            \eqno(60)$$
    defined on the space $\oQ = \wQ \times \R_+$, where $\wQ$ is the tangent bundle $\sT M$ with the image of the zero
section removed is treated as a family
        $$L(\dot x,\,\cdot\,) \,\colon \R_+ \rightarrow \R \,\colon \zm \mapsto {{1}\over{2\zm}} \langle g(\dot q), \dot
q\rangle
                                                                                                            \eqno(61)$$
    of functions on the fibres of the projection
        $$\zh \,\colon \oQ \rightarrow \wQ.
                                                                                                            \eqno(62)$$
    The dynamics is the set
        $$\eqalign{
    D = \left\{\VPU{14pt}\dot p \in \sT\sT^\*M;\;\right.& \sT\zp_M(\dot p) \in \wQ, \;\;\langle g(\sT\zp_M(\dot p)),
\sT\zp_M(\dot p)\rangle = 0, \cr
    & \left.\exi{\zm \in \R_+} \zt_{\sT^*M}(\dot p) = {{1}\over{\zm}}g(\sT\zp_M(\dot p)),\;\;\dot p \in
H_{\zt_{\sT^*M}(\dot p)} \right\}
    }
                                                                                                            \eqno(63)$$
        }

        \Example{
    The Hamiltonian generating object for the dynamics of Example 3 is the function
        $$H \,\colon \oP \rightarrow \R \,\colon (p,\zm) \mapsto {{\zm}\over{2}}\langle p, g^{-1}(p)\rangle
                                                                                                            \eqno(64)$$
    defined on $\oP = \wP \times \R_+$, where $\wP$ is the cotangent bundle $\sT^\*M$ with the image of the zero section
removed is treated as a family of functions on fibres of the projection
        $$\zz \,\colon \oP \rightarrow \wP \,\colon (p,\zm) \mapsto p.
                                                                                                            \eqno(65)$$
        }

    It is obvious that the set $S$ is a Lagrangian submanifold if the first two operations listed above produce a
Lagrangian submanifold.  For this reason we will concentrate our attention on simpler generating objects with trivial Liouville structures and
unconstrained families of functions.  Such simple generating objects are encountered in the theory of partially controlled static systems.  Variational
formulations of dynamics require the use of nontrivial Liouville structures as is seen in the above example.  We will derive conditions sufficient for
obtaining Lagrangian submanifolds from the simple generating ojects.

        \sect{Families of functions and sets generated by families.}
    The diagram
    \vskip1mm
        $$\vcenter{
    \begindc{0}[1]
    \obj(000,055)[01]{$\oQ$}
    \obj(060,055)[11]{$\R$}
    \obj(000,000)[00]{$Q$}
    \obj(060,000)[10]{$$}
    \mor{01}{11}[10,10]{$\oU$}[1,0]
    \mor{01}{00}[10,10]{$\zh$}[0,0]
    \enddc}
                                                                                                            \eqno(66)$$
    \vskip2mm
    \noindent representing a simple generating object is relevant for our analysis.  This simple object can be obtained
from the diagram \RF(28) by setting $\wQ = Q$ and idenifying the symplectic space $(P,\zw)$ with $(\sT^\*Q,\zw_Q)$ or it can be considered an essential
portion of the complete diagram \RF(28).

    The set
        $$\sC\sr(\oU,\zh) = \left\{\oq \in \oQ ;\; \all{\zd\oq \in \sV_{\oq} \oQ} \langle \rd \oU, \zd\oq \rangle = 0
\right\}
                                                                                                            \eqno(67)$$
    is the {\it critical set} of the family $(\oU,\zh)$.  Elements of the critical set are {\it critical points} of
$(\oU,\zh)$.  There is a mapping $\zk(\oU,\zh) \,\colon \sC\sr(\oU,\zh) \rightarrow \sT^\*Q$ characterized by
        $$\langle \zk(\oU,\zh)(\oq), \zd\tq \rangle = \langle \rd \oU, \zd\oq \rangle
                                                                                                            \eqno(68)$$
    for each $\zd\tq \in \sT_{\zh(\oq)}Q$ and each $\zd\oq \in \sT_\oq \oQ$ such that $\sT\zh(\zd\oq) = \zd\tq$.

    The family of functions $(\oU,\zh)$ generates a set $S \in \sT^\*Q$.  This set is obtained by one of the two
following constructions.
        \List
    \Item{1} Let $\oS = \im(\rd \oU) \subset \sT^\*\oQ$ be the Lagrangian submanifold generated by the function $\oU$.
The symplectic relation $\sPh\,\zh$ applied to $\oS$ produces the set
        $$\sPh\,\zh(\oS) = \t\zh(\oS \cap \sV^\polar \oQ).
                                                                                                            \eqno(69)$$
    This is the set $S$ generated by the family $(\oU,\zh)$.

    \Item{2} The set $S$ is the image of $\zk(\oU,\zh)$.  The formula
        $$S = \left\{\tf \in \sT^\*Q \;;\; \exi{\oq \in \oQ} \zh(\oq) = \zp_Q(\tf) \;\;\all{\zd\oq \in \sT_\oq\oQ}
\langle \rd\oU, \zd\oq\rangle = \langle \tf,\, \sT\zh(\zd\oq)\rangle \right\}
                                                                                                            \eqno(70)$$
    gives an explicit description.
        \endList

        \sect{Examples.}
    We give examples of constitutive sets of static systems derived from variational principles applied to families of
functions.  Variational principles of statics are models for all variational principles of classical physics since at the basis of a variational
principle there is a Liouville structure formally identifying the principle with that of a static system.  Configuration spaces will be constructed
using an affine space $Q$.  The model space is a vector space $V$ of dimension 3 with a Euclidean metric $g \,\colon V \rightarrow V^\*$.

        \Example{
    A material point with configuration $q_2$ in the affine space $Q$ is connected to a fixed point $q_0$ with a rigid
rod of length $a$.  A second material point with configuration $q_1$ is tied elastically to $q_2$ with a spring of spring constant $k$.  The internal
configuration space $\oQ$ is the product $Q \times D$, with
        $$D = \left\{q_2 \in Q ;\; \|q_2 - q_0\| = a \right\}.
                                                                                                            \eqno(71)$$
    The set
        $$\sT D = \left\{(q_2,\zd q_2) \in Q \times V ;\; \|q_2 - q_0\| = a,\; \langle g(q_2 - q_0), \zd q_2\rangle = 0
\right\}
                                                                                                            \eqno(72)$$
    is the tangent bundle of $D$ and the set
        $$\sT^\*D = \left\{(q_2,f_2) \in D \times V^\* \;;\; \langle f_2, q_2 - q_0\rangle = 0 \right\}
                                                                                                            \eqno(73)$$
    is chosen to represent the dual of $\sT D$.  We have the identifications
        $$\sT\oQ = Q \times V \times \sT D
                                                                                                            \eqno(74)$$
        and
        $$\sT^\*\oQ = Q \times V^\* \times \sT^\*D.
                                                                                                            \eqno(75)$$
    \dacapo
    The internal energy
        $$\oU \,\colon \oQ \rightarrow \R \,\colon (q_1,q_2) \mapsto {{k}\over{2}}\langle g(q_2 - q_1), q_2 - q_1\rangle
                                                                                                            \eqno(76)$$
    of the system generates the {\it internal constitutive set}
        $$\eqalign{ \oS &= \left\{(q_1,f_1,q_2,f_2) \in Q \times V^\* \times \sT^\*D ;\; \; f_1 = kg(q_1 - q_2), \right.
    \hskip10mm\cr
    &\hskip30mm \left.  f_2 - kg(q_2 - q_1) = a^{-2}\langle f_2 - kg(q_2 - q_1), q_2 - q_0\rangle g(q_2 - q_0) \right\}.
    }
                                                                                                            \eqno(77)$$
    This set is the image of the differential $\rd\oU$.
    \dacapo
    The configuration $q_2$ is not controlled.  The control configuration space is the space $Q$.  The projection
        $$\zh \,\colon \oQ \rightarrow Q \,\colon (q_1,q_2) \mapsto q_1
                                                                                                            \eqno(78)$$
    is the {\it control relation}.  The set
        $$\sV\oQ = \left\{(q_1,\zd q_1,q_2,\zd q_2) \in Q \times V \times \sT D;\; \zd q_1 = 0 \right\}
                                                                                                            \eqno(79)$$
    is the vertical bundle and the set
        $$\sV^\polar\oQ = \left\{(q_1,f_1,q_2,f_2) \in Q \times V^\* \times \sT^\*D;\; f_2 = 0 \right\}
                                                                                                            \eqno(80)$$
    is its polar.  The strict symplectic reduction is the mapping
        $$\t\zh \,\colon \sV^\polar\oQ \rightarrow Q \times V^\* \,\colon (q_1,f_1,q_2,f_2) \mapsto (q_1,f_1).
                                                                                                            \eqno(81)$$
    The set
        $$\left\{(q_1,f_1,q_2,f_2) \in Q \times V^\* \times \sT^\*D ;\; f_1 = kg(q_1 - q_2), f_2 = 0, \|q_1 - q_0\|(q_2
- q_0) = \pm(q_1 - q_0) \right\}
                                                                                                            \eqno(82)$$
    is the intersection $\oS \cap \sV^\polar\oQ$.  The {\it constitutive set}
        $$\eqalign{
    S &= \left\{(q_1,f_1) \in Q \times V^\* ;\; \|f_1\| = ka \;\;\;{\rm if }\;\; \; q_1 = q_0, \vphantom{\|^{-1}}\right.
    \cr
    &\hskip30mm \left.  f_1 = k\left(1 \pm a\|q_1 - q_0\|^{-1}\right)g(q_1 - q_0) \;\;\;{\rm if }\;\; \; q_1 \neq q_0
\right\}.
    }
                                                                                                            \eqno(83)$$
    of the partially controlled system is obtained from $\oS$ by applying the symplectic reduction relation $\sPh\,\zh$.
    It is the image of $\oS \cap \sV^\polar\oQ$ by the mapping $\t\zh$.
    \dacapo
    The internal energy is treated as a family of functions $(\oU,\zh)$ defined on fibres of the projection $\zh$.  The
critical set
        $$\eqalign{
    \sC\sr(\oU,\zh) &= \left\{(q_1,q_2) \in \oQ ;\; \|q_1 - q_0\|(q_2 - q_0) = \pm a(q_1 - q_0) \right\} \cr
    &= \left\{(q_1,q_2) \in \oQ ;\; (q_2 - q_1) = a^{-2}\langle g(q_2 - q_1), q_2 - q_0\rangle (q_2 - q_0)\right\} \cr
    }
                                                                                                            \eqno(84)$$
    is a submanifold of $\oQ$.  This observation will be confirmed subsequently.  The constitutive set $S$ is the image
of the injective mapping
        $$\zk(\oU,\zh) \,\colon \sC\sr(\oU,\zh) \rightarrow Q \times V^\* \,\colon (q_1,q_2) \mapsto (q_1, kg(q_1 -
q_2)).
                                                                                                            \eqno(85)$$
    The constitutive set can be obtained directly from the variational definition
        $$S = \left\{(q_1,f_1) \in Q \times V^\* ;\; \exi{q_2 \in D} \; \all{\zd q_1 \in \sT Q,\; \zd q_2 \in \sT D}
\;\; k\langle g(q_2 - q_1), \zd q_2 - \zd q_1\rangle = \langle f_1, \zd q_1\rangle \right\}.
                                                                                                            \eqno(86)$$
    \dacapo
    We show that $S$ is a submanifold of $\sT^\*Q$.  With the exclusion of the set
        $$\left\{(q_1,f_1) \in Q \times V^\* ;\; q_1 = q_0,\;\;\|f_1\| = ka \right\}
                                                                                                            \eqno(87)$$
    the set $S$ is the union of images of the two smooth sections
        $$\zs^+ \,\colon Q\setminus \{q_0\} \rightarrow Q \times V^\* \,\colon q_1 \mapsto \left(q_1, 1 + a\|q_1 -
q_0\|^{-1}g(q_1 - q_0)\right)
                                                                                                            \eqno(88)$$
        and
        $$\zs^- \,\colon Q\setminus \{q_0\} \rightarrow Q \times V^\* \,\colon q_1 \mapsto \left(q_1, 1 - a\|q_1 -
q_0\|^{-1}g(q_1 - q_0)\right).
                                                                                                            \eqno(89)$$
    The set
        $$\left\{(q_1,f_1) \in Q \times V^\* ;\; g(q_1 - q_0) + \|f_1\|^{-1}(a - k^{-1}\|f_1\|)f_1 = 0 \right\}
                                                                                                            \eqno(90)$$
    is the set $S$ with the exclusion of
        $$\left\{(q_1,f_1) \in Q \times V^\* ;\; \|q_1 - q_0\| \geqs\, a \right\}.
                                                                                                            \eqno(91)$$
    The set \RF(90) is the image of the smooth section
        $$\zr \,\colon V^\* \rightarrow Q \times V^\* \,\colon f_1 \mapsto \left(\|f_1\|^{-1}(k^{-1}\|f_1\| -
a)g^{-1}(f_1), f_1\right)
                                                                                                            \eqno(92)$$
    of the canonical projection of $Q \times V^\*$ onto $V^\*$.  It follows that $S$ is a submanifold of $Q \times V^\*$
of dimension 3.
    }

        \Example{
    A material point with configuration $q_1$ in the affine space $Q$ is tied elastically to a fixed point $q_0$ with a
spring of spring constant $k_1$.  A second material point with configuration $q_2$ is tied elastically to $q_1$ with a spring of spring constant $k_2$
and rest length $a$.  The internal configuration space $\oQ$ is the product $Q \times Q$ and the internal energy is the function
        $$\oU \,\colon Q \times Q \rightarrow \R \,\colon (q_1,q_2) \mapsto {{k_1}\over{2}}\langle g(q_1 - q_0), q_1 -
q_0\rangle + {{k_2}\over{2}}\left(\sqrt{\langle g(q_2 - q_1), q_2 - q_1\rangle} - a\right)^2.
                                                                                                            \eqno(93)$$
    The internal energy generates the internal constitutive set
        $$\eqalign{
    \oS &= \left\{(q_1,f_1,q_2,f_2) \in Q \times V^\* \times Q \times V^\* ;\; f_1 + f_2 = k_1g(q_1 - q_0) , \VPU{15pt}
\right.  \hskip20mm\cr
    &\hskip57mm \left.  f_2 = k_2\left(1 - {{a}\over{\|q_1 - q_2\|}}\right)g(q_2 - q_1) \right\}.
    }
                                                                                                            \eqno(94)$$
    \dacapo
    The configuration $q_2$ is not controlled.  The control configuration space is the space $Q$.  The projection
        $$\zh \,\colon \oQ \rightarrow Q \,\colon (q_1,q_2) \mapsto q_1
                                                                                                            \eqno(95)$$
    is the {\it control relation}.  The set
        $$\sV\oQ = \left\{(q_1,\zd q_1,q_2,\zd q_2) \in Q \times V \times Q \times V ;\; \zd q_1 = 0 \right\}
                                                                                                            \eqno(96)$$
    is the vertical bundle and the set
        $$\sV^\polar\oQ = \left\{(q_1,f_1,q_2,f_2) \in Q \times V^\* \times Q \times V^\*;\; f_2 = 0 \right\}
                                                                                                            \eqno(97)$$
    is its polar.  The strict symplectic reduction is the mapping
        $$\t\zh \,\colon \sV^\polar\oQ \rightarrow Q \times V^\* \,\colon (q_1,f_1,q_2,f_2) \mapsto (q_1,f_1).
                                                                                                            \eqno(98)$$
    The intersection $\oS \cap \sV^\polar\oQ$ is the set
        $$\left\{(q_1,f_1,q_2,f_2) \in Q \times V^\*Q \times Q \times V^\* ;\; \|q_2 - q_1\| = a, f_1 = k_1 g(q_1 -
q_0), f_2 = 0 \right\}.
                                                                                                            \eqno(99)$$
    The {\it constitutive set}
        $$S = \left\{(q_1,f_1) \in Q \times V^\* ;\; f_1 = k_1 g(q_1 - q_0) \right\}.
                                                                                                           \eqno(100)$$
    of the partially controlled system is obtained from $\oS$ by applying the symplectic reduction relation $\sPh\,\zh$.
    It is the image of $\oS \cap \sV^\polar\oQ$ by the mapping $\t\zh$.  This constitutive set is the image of the
mapping
        $$\zk(\oU,\zh) \,\colon \sC\sr(\oU,\zh) \rightarrow Q \times V^\* \,\colon (q_1,q_2) \mapsto (q_1, kg(q_1 -
q_0)),
                                                                                                           \eqno(101)$$
    defined on the critical set
        $$\sC\sr(\oU,\zh) = \left\{(q_1,q_2) \in Q \times Q ;\; \|q_2 - q_1\| = a \right\}.
                                                                                                           \eqno(102)$$
    The constitutive set can also be obtained from the variational construction
        $$S = \left\{(q_1,f_1) \in Q \times V^\* ;\; \exi{q_2 \in D} \; \all{\zd q_1 \in \sT Q,\; \zd q_2 \in \sT D}
\;\; k\langle g(q_2 - q_1), \zd q_2 - \zd q_1\rangle = \langle f_1, \zd q_1\rangle \right\}.
                                                                                                           \eqno(103)$$
    }

        \sect{Regular reductions of Lagrangian submanifolds.}
        Let
        $$\t\zh \,\colon \oN \rightarrow \sT^\*Q
                                                                                                           \eqno(104)$$
    be a strict symplectic reduction from a coistropic submanifold $\oN$ of a symplectic manifold $(\sT^\*\oQ,\zw_\oQ)$
onto a symplectic manifold $(\sT^\*Q,\zw_Q)$ and let $\oS$ be a Lagrangian submanifold of the symplectic manifold $(\sT^\*\oQ,\zw_\oQ)$.  We are
extracting from [1] and [5] the following facts about the reduced set $S = \t\zh(\oS)$.  We assume that the intersection of $\oS$ with $\oN$ is not
empty.

        \List
    \Item{1)} If the intersection of $\oS$ with $\oN$ is clean, then $S$ is an immersed Lagrangian submanifold of
$(\sT^\*Q,\zw_Q)$.
    \Item{2)} If $\oS$ is transverse to $\oN$, then $S$ is an immersed Lagrangian submanifold of $(\sT^\*Q,\zw_Q)$ and
$\t\zh|(\oN \cap \oS)$ is an immersion.
        \endList

    Recall that submanifolds $\oS$ and $\oN$ have {\it clean intersection} if $\oS \cap \oN \subset \sT^\*\oQ$ is a
submanifold and
        $$\sT_\of(\oS \cap \oN) = \sT_\of\oS \cap \sT_\of\oN
                                                                                                           \eqno(105)$$
    at each $\of \in \oS \cap \oN$.  The submanifold $\oS$ is {\it transverse} to $\oN$ if
        $$\sT_\of\oS + \sT_\of\oN = \sT_\of\sT^\*\oQ.
                                                                                                           \eqno(106)$$
        \Example{
    We use the notation of Example~5.  Let $\bf = (q_1, f_1, q_2, f_2) \in \oS \cap \sV^\comp \oQ$, i.e.
        $$f_2=0,\ f_1= kg(q_1 - q_2), \ g(q_2 - q_1)= a^{-2} \langle g(q_2 - q_1), q_2 - q_0 \rangle g(q_2 - q_0), \
\|q_2 - q_0\| = a.
                                                                                                           \eqno(107)$$
    We have
        $$\sT_\bif\sT^\* \oQ = \left\{(\zd q_1, \zd f_1, \zd q_2, \zd f_2) \in V\times V^\* \times V \times V^\* ;\;
    \langle g(q_2 - q_0),\zd q_2 \rangle = 0, \ \langle \zd f_2, q_2 - q_0 \rangle = 0 \right\},
                                                                                                           \eqno(108)$$
        $$\sT_\bif \oN = \left\{\zd q_1, \zd f_1, \zd q_2, \zd f_2) \in V\times V^\* \times V \times V^\* ;\; \langle
g(q_2 - q_0),\zd q_2 \rangle = 0, \ \zd f_2 = 0 \right\},
                                                                                                           \eqno(109)$$
    and
        $$\eqalign{
    \sT_\bif \oS &= \left\{(\zd q_1,\zd f_1,\zd q_2,\zd f_2) \in V \times V^\* \times V \times V^\* ;\; \langle g(q_2 -
q_0),\zd q_2 \rangle = 0, \ \langle \zd f_2, q_2 - q_0, \rangle = 0, \right.  \cr
    &\hskip7mm \left.  \zd f_1 = kg(\zd q_1 - \zd q_2), \ \zd f_2 = kg(\zd q_2 - \zd q_1) - a^{-2} \langle kg(\zd q_2 -
\zd q_1),q_2 -q_0 \rangle g(q_2 - q_0) \right.  \cr
    &\hskip7mm \left.  -a^{-2}\langle kg(q_2 -q_1), \zd q_2 \rangle g(q_2 - q_0) - a^{-2}\langle kg(q_2 -q_1), q_2 - q_0
\rangle g(\zd q_2)\VPU{15pt}  \right\}
    }
                                                                                                           \eqno(110)$$
    For every $\zd \bif = (\zd q_1,\zd f_1,\zd q_2,\zd f_2)\in \sT_\bif \sT^\*\oQ$,
    we put
        $$\zd_1 \bif = (-{{1}\over{k}}g ^{-1}(\zd f_2), -\zd f_2, 0, \zd f_2)
                                                                                                           \eqno(111)$$
        and
        $$ \zd_2 \bif = (\zd q_1 - {{1}\over{k}}g ^{-1}(\zd f_2), \zd q_2, \zd f_1 + \zd f_2, 0).
                                                                                                           \eqno(112)$$
    A direct check shows that
    $\zd_1 \bif \in \sT_\bif \oS $, $\zd_2\bif \in \sT_\bif \oN$.  Since $\zd_1 \bif + \zd_2 \bif = \zd \bif$, we have
        $$ \sT_\bif \sT^\*\oQ = \sT_\bif \oN + \sT_\bif \oS.
                                                                                                           \eqno(113)$$
     We conclude that $\oS$ is transverse to $\oN$ at $\bif$.
        }

        \Example{
    We use the notation of Example~6.  Let $\bif = (q_1, f_1, q_2, f_2) \in \oS \cap \sV^\comp \oQ$, i.e.  $\|q_1 - q_2\|
= a$, $f_1= k_1 g(q_1 -q_0)$, and $f_2 = 0$.  We have $\sT_\bif \sT^\* \oQ = V\times V^\*\times V \times V^\*$,
        $$ \sT_\bif\sV^\comp \oQ = \left\{(\zd q_1, \zd f_1, \zd q_2, \zd f_2) \in V\times V^\* \times V \times V^\* ;\;
f_2 = 0 \right\},
                                                                                                           \eqno(114)$$
    and
        $$\eqalign{\sT_\bif \oS &= \left\{(\zd q_1, \zd f_1, \zd q_2, \zd f_2) \in V\times V^\* \times V \times V^\* ;\;
\zd f_1 + \zd f_2 = k_1g(\zd q_1), \right. \cr
    &\hskip7mm \left. \zd f_2 = -k_2 {{a}\over{\|q_1 - q_2\|^3}}\langle g(q_1 - q_2),\zd q_1 - \zd q_2
\rangle g(q_2 - q_1) \right\}.
    }
                                                                                                           \eqno(115)$$
    Since $\zd f_2$ is proportional to $g(q_2 - q_1)$, the algebraic sum $\sT_\bif \sV^\comp \oQ + \sT_\bif \oS$ is not,
for $\dim \sV>1$, equal to $\sT_\bif \sT^\*\oQ$ and  $\oS$ is not transverse to $\sV^\comp \oQ$.  On the other hand,
        $$ \oS \cap \sV^\comp \oQ = \left\{(q_1, f_1, q_2, f_2) \in V \times V^\* \times V \times V^\* ;\; \|q_2 - q_1\|
= a,\ f_1 = k_1 g(q_1 - q_0), \ f_2 = 0\right\}
                                                                                                           \eqno(116)$$
    is a submanifold, and
        $$\eqalign{\sT_\bif(\oS \cap \sV^\comp \oQ) &= \left\{(\zd q_1, \zd f_1, \zd q_2, \zd f_2) \in V\times V^\*
\times V \times V^\* ;\; \langle g(q_2 - q_1),\zd q_2 - \zd q_1 \rangle = 0, \right.  \cr
    &\hskip7mm \left. \zd f_1 = k_1g(\zd q_1), \  \zd f_2 = 0 \right\}.
    }
                                                                                                           \eqno(117)$$
    Comparing \RF(117) with \RF(114) and \RF(115), we establish the equality
        $$\sT_\bif(\oS \cap \sV^\comp \oQ) = \sT_\bif(\oS)\cap \sT_\bif(\sV^\comp \oQ).
                                                                                                           \eqno(118)$$
    It follows that $\oS$ and $\sV^\comp \oQ$ have clean intersection.
        }

        \sect{The Hessian of a function at a critical point.}
    Let $Q$ be a differential manifold and let $q$ be a critical point of a function
        $$U \,\colon Q \rightarrow \R.
                                                                                                           \eqno(119)$$
    The image of the differential
        $$\rd U \,\colon Q \rightarrow \sT^\*Q
                                                                                                           \eqno(120)$$
    is a Lagrangian submanifold $S \subset \sT^\*Q$ of the symplectic space $(\sT^\*Q,\zw_Q)$ .  It intersects the image
of the zero section
        $$O_{\zp_Q} \,\colon Q \rightarrow \sT^\*Q
                                                                                                           \eqno(121)$$
    at $f = \rd U(q) = O_{\zp_Q}(q)$.  The tangent space $S_f = \sT_f S = \sT\rd U(\sT_q Q)$ is a Lagrangian subspace of
the symplectic vector space $(\sT_f\sT^\*Q,\zw_f)$.  We use the decomposition
        $$H_f + V_f
                                                                                                           \eqno(122)$$
    of the space $\sT_f\sT^\*Q$ introduced in Section 2.  The function $F = 0$ is used.  It follows that $\rd F =
O_{\zp_Q}$.  The decomposition makes it possible to define a quadratic generating function
        $$h \,\colon \sT_q Q \rightarrow \R \,\colon \zd q \mapsto {{1}\over{2}} \langle p_v(\sT\rd U(\zd q)), \zd
q\rangle.
                                                                                                           \eqno(123)$$
    The {\it Hessian} of $U$ at the critical point $q$ is the bilinear symmetric function
        $$\sH(U,q) \,\colon \sT_q Q \times \sT_q Q \rightarrow \R
                                                                                                           \eqno(124)$$
    defined as the {\it polarization}
        $$\zd h \,\colon \sT_q Q \times \sT_q Q \rightarrow \R \,\colon (\zd_1 q,\zd_2 q) \mapsto h(\zd_1 q + \zd_2 q) -
h(\zd_1 q) - h(\zd_2 q)
                                                                                                           \eqno(125)$$
    of the quadratic function $h$.  It follows from elementary linear symplectic algebra that the function $h$ is
quadratic and its polarization is a symmetric bilinear mapping.  The space $S_f$ is generated by $h$ in the sense that
        $$S_f = \left\{\zd f \in \sT_f\sT^\*Q ;\; \all{\zd q \in \sT_q Q}\;\langle p_v(\zd f), \zd q\rangle = \zd
h(p_h(\zd f),\zd q) \right\}.
                                                                                                           \eqno(126)$$
    It follows from this expression for $S_f = \sT\rd U(\sT_q Q)$ that
        $$\sH(U,q)(\zd_1 q,\zd_2 q) = \langle p_v(\sT\rd U(\zd_1 q)), \zd_2 q\rangle.
                                                                                                           \eqno(127)$$
    A useful expression
        $$\sH(U,q)(\zd_1 q,\zd_2 q) = \zw_f(\sT\rd U(\zd_1 q),\sT O_{\zp_Q}(\zd_2 q)) = \langle\zw_Q, \sT\rd U(\zd_1 q)
\wedge \sT O_{\zp_Q}(\zd_2 q)\rangle
                                                                                                           \eqno(128)$$
    is derived by using the formula \RF(25).

    The image
        $$\zC(S_f) \subset \sT_q Q \oplus \sT_q^\*Q
                                                                                                           \eqno(129)$$
    is a Lagrangian subspace denoted by $L_f$.  This subspace is the graph of the linear mapping
        $$\zl_f \,\colon \sT_q Q \rightarrow \sT_q^\*Q \,\colon \zd q \mapsto p_v(\sT\rd U(\zd q))
                                                                                                           \eqno(130)$$
    symmetric in the sense that
        $$\langle \zl_f(\zd_1 q), \zd_2 q\rangle = \langle \zl_f(\zd_2 q), \zd_1 q\rangle.
                                                                                                           \eqno(131)$$
    For the Hessian we have the expression
        $$\sH(U,q)(\zd_1 q,\zd_2 q) = \langle \zl_f(\zd_1 q), \zd_2 q\rangle.
                                                                                                           \eqno(132)$$

    In the following two propositions we are using a critical point $q$ of a function $U\,\colon Q \rightarrow \R$,
vectors $\zd_1 q$ and $\zd_2 q$ in $\sT_q Q$, and a choice of a mapping $\zq \,\colon \R^2 \rightarrow Q$ such that $\zq(0,0) = q$, $\st\zq(\cdot,0)(0)
= \zd_1 q$, and $\st\zq(0,\cdot)(0) = \zd_2 q$.

        \Proposition{
    The derivative
        $$\rD^{(1,1)}(U \comp \zq)(0,0)
                                                                                                           \eqno(133)$$
    of a function $U \,\colon Q \rightarrow \R$ depends on $\zd_1 q$ and $\zd_2 q$ but not on the choice of the mapping
$\zq$.  }
        \Proof{
        $$\rD^{(1,1)}(U \comp \zq)(0,0) = \rD^{(1,1)}((U - U(q)1) \comp \zq)(0,0)
                                                                                                           \eqno(134)$$
    and $U - U(q)1$ is in $\sI_1(Q,q) = (\sI_0(Q,q))^2$. $\sI_0(Q,q)$ is the maximal ideal of functions related to $q$.
     It is sufficient to examine the expression \RF(133) for $U = FG$ with $F$ and $G$ in $I_0(Q,q)$.  The equality
        $$\eqalign{
    \rD^{(1,1)}(FG \comp \zq)(0,0) &= \rD^{(1,1)}((F \comp \zq)(G \comp \zq))(0,0) \cr
    &= \rD^{(0,1)}\left(\rD^{(1,0)}(F \comp \zq)\rD^{(0,0)}(G \comp \zq)\right.  \cr
    &\hskip10mm + \left.\rD^{(0,0)}(F \comp \zq)\rD^{(1,0)}(G \comp \zq)\right)(0,0) \cr
    &= \left(\rD^{(1,1)}(F \comp \zq)\rD^{(0,0)}(G \comp \zq) + \rD^{(1,0)}(F \comp \zq)\rD^{(0,1)}(G \comp \zq)\right.
\cr
    &\hskip10mm + \left.\rD^{(0,1)}(F \comp \zq)\rD^{(1,0)}(G \comp \zq) + \rD^{(0,0)}(F \comp \zq)\rD^{(1,1)}(G \comp
\zq)\right)(0,0) \cr
    &= \left(\rD^{(1,0)}(F \comp \zq)\rD^{(0,1)}(G \comp \zq) + \rD^{(0,1)}(F \comp \zq)\rD^{(1,0)}(G \comp \zq)
\right)(0,0) \cr
    &= \langle \rd F, \zd_1 q \rangle \langle \rd G, \zd_2 q \rangle + \langle \rd F, \zd_2 q \rangle \langle \rd G,
\zd_1 q \rangle
    }
                                                                                                           \eqno(135)$$
    proves the proposition.
    }

        \Proposition{
    The Hessian $\sH(U,q)$ is the bilinear symmetric mapping
        $$(\zd_1 q,\zd_2 q) \mapsto \rD^{(1,1)}(U \comp \zq)(0,0).
                                                                                                           \eqno(136)$$
    }
        \Proof{
    We choose a mapping $\zc\,\colon \R^2 \rightarrow \sT^\*Q$ such that $\zp_Q \comp \zc = \zq$, $\zc(\cdot,0) = \rd U
\comp \zq(\cdot,0) $, and $\zc(0,\cdot) = O_{\zp_Q} \comp \zq(0,\cdot))$.  The mapping $\zc$ represents the pair
        $$(\sT\rd U(\zd_1 q),\sT O_{\zp_Q}(\zd_2 q)) \in \sT_f\sT^\*Q \times \sT_f\sT^\*Q
                                                                                                           \eqno(137)$$
    since
        $$\sT\rd U(\zd_1 q) = \sT\rd U(\st\zq(\cdot,0)(0)) = \st(\rd U \comp \zq(\cdot,0))(0) = \st\zc(\cdot,0)(0)
                                                                                                           \eqno(138)$$
        and
        $$\sT O_{\zp_Q}(\zd_2 q) = \sT O_{\zp_Q}(\st\zq(0,\cdot)(0)) = \st(O_{\zp_Q} \comp \zq(0,\cdot))(0) =
\st\zc(0,\cdot)(0).
                                                                                                           \eqno(139)$$
    The equality
        $$\eqalign{
    \sH(U,q)(\zd_1 q,\zd_2 q)&= \langle \zw_Q, \sT\rd U(\zd_1 q) \wedge \sT O_{\zp_Q}(\zd_2 q)\rangle \cr
    &= {{\rd}\over{\rd s}}\langle \zy_Q, \st\zc(s,\cdot\,)(0)\rangle\big|_{s=0} - {{\rd}\over{\rd s}}\langle \zy_Q,
\st\zc(\cdot,s)(0)\rangle\big|_{s=0} \cr
    &= {{\rd}\over{\rd s}}\langle \zt_{\sT^\*Q}(\st\zc(s,\cdot\,)(0)), \sT\zp_Q(\st\zc(s,\cdot\,)(0))\rangle\big|_{s=0}
\cr
    & \hskip37mm - {{\rd}\over{\rd s}}\langle \zt_{\sT^\*Q}(\st\zc(\,\cdot,s)(0)),
\sT\zp_Q(\st\zc(\,\cdot,s)(0))\rangle\big|_{s=0} \cr
    &= {{\rd}\over{\rd s}}\langle \zc(s,0), \st(\zp_Q \comp \zc)(s,\cdot\,)(0)\rangle\big|_{s=0} - {{\rd}\over{\rd
s}}\langle \zc(0,s), \st(\zp_Q \comp \zc)(\,\cdot,s)(0)\rangle\big|_{s=0} \cr
    &= {{\rd}\over{\rd s}}\langle(\rd U \comp \zq)(s,0), \st\zq(s,\cdot\,)(0)\rangle\big|_{s=0} - {{\rd}\over{\rd
s}}\langle(O_\zp \comp \zc)(0,s), \st\zq(\,\cdot,s)(0)\rangle\big|_{s=0} \cr
    &= {{\rd}\over{\rd s}}\langle(\rd U(\zq(s,0)), \st\zq(s,\cdot\,)(0)\rangle\big|_{s=0} \cr
    &= {{\partial}\over{\partial s}}{{\partial}\over{\partial t}}U(\zq(s,t))\big|_{s=0,\, t=0} \cr
    &= \rD^{(1,1)}(U \comp \zq)(0,0).
    }
                                                                                                           \eqno(140)$$
    proves the proposition.  }

    The last proposition offers an alternate definition of the Hessian.  This definition is closer to the usual
definition of the Hessian in terms of local coordinates.

    If $q$ is not a critical point of the function $U$, then a Hessian of $U$ at $q$ can be defined in relation to a
function $F$ on $Q$ such that $\rd F(q) = \rd U(q)$.  This {\it relative Hessian} is the Hessian $\sH(U - F,q)$.

        \sect{The Hessian of a family of functions at a critical point.}
    If $\oq \in \oQ$ is a critical point of a family

    \vskip1mm
        $$\vcenter{
    \begindc{0}[1]
    \obj(000,055)[01]{$\oQ$}
    \obj(060,055)[11]{$\R$}
    \obj(000,000)[00]{$Q$}
    \obj(060,000)[10]{$$}
    \mor{01}{11}[10,10]{$\oU$}[1,0]
    \mor{01}{00}[10,10]{$\zh$}[0,0]
    \enddc}
                                                                                                           \eqno(141)$$
    \vskip2mm
    \noindent then $\rd\oU(\oq)$ is in $\sV_\oq^\polar Q$.  It follows that
        $$\langle \rd\oU(\oq), \zd\oq\rangle = \langle \t\zh(\rd\oU(\oq)), \sT\zh(\zd\oq)\rangle
                                                                                                           \eqno(142)$$
    for each $\zd\oq \in \sT_\oq \oQ$.  Let $F$ be a function on $Q$ such that $\rd F(\zh(\oq)) = \t\zh(\rd\oU(\oq))$
and let $\oF = F \comp \zh$.  For each $\zd\oq \in \sT_\oq \oQ$, we have
        $$\langle \rd\oF(\oq), \zd\oq\rangle = \langle (\zh^\*\rd F)(\oq), \zd\oq\rangle = \langle \rd F(\zh(\oq)),
\sT\zh(\zd\oq)\rangle.
                                                                                                           \eqno(143)$$
    Hence, $\rd\oF(\oq) = \rd\oU(\oq)$.  We examine the bilinear mapping
        $$\sV_\oq\oQ \times \sT_\oq\oQ \rightarrow \R \,\colon (\zd_1\oq,\zd_2\oq) \mapsto \sH(\oU -
\oF,\oq)(\zd_1\oq,\zd_2\oq)
                                                                                                           \eqno(144)$$
    extracted from the relative Hessian
        $$\sH(\oU - \oF,\oq) \,\colon \sT_\oq\oQ \times \sT_\oq\oQ \rightarrow \R.
                                                                                                           \eqno(145)$$
    The mapping
        $$\zq \,\colon \R^2 \rightarrow \oQ
                                                                                                           \eqno(146)$$
    representing a pair $(\zd_1\oq,\zd_2\oq) \in \sV_\oq\oQ \times \sT_\oq\oQ$ can be chosen to be vertical in the sense
that
        $$(\zh \comp \zq)(s_1,s_2) = (\zh \comp \zq)(0,s_2).
                                                                                                           \eqno(147)$$
    For the function $\oF$ we have
        $$(\oF \comp \zq)(s_1,s_2) = (F \comp \zh \comp \zq)(s_1,s_2) = (F \comp \zh \comp \zq)(0,s_2) = (\oF \comp
\zq)(0,s_2).
                                                                                                           \eqno(148)$$
    It follows that
        $$\sH(\oU - \oF,\oq)(\zd_1\oq,\zd_2\oq) = \rD^{(1,1)}((\oU \comp \zq) - (\oF \comp \zq))(0,0) = \rD^{(1,1)}(\oU
\comp \zq)(0,0).
                                                                                                           \eqno(149)$$
    We had to choose a function $F$ to be able to define the relative Hessian $\sH(\oU - \oF,\oq)$.  It turns out that
the choice of this function has no effect on the construction of the mapping \RF(144).  We define the {\it Hessian} of the family \RF(141) at the
critical point $\oq$ as the bilinear mapping
        $$\sH(\oU,\zh,\oq) \,\colon \sV_\oq Q \times \sT_\oq Q \rightarrow \R \,\colon (\zd_1\oq,\zd_2\oq) \mapsto
\sH(\oU - \oF,\oq)(\zd_1\oq,\zd_2\oq).
                                                                                                           \eqno(150)$$

        \Example{
    We consider the generating family of Example~5.  Let $\oq = (q_1,q_2) \in \sC\sr(\oU,\zh))$, $\zd_2\oq = (\zd_2 q_1,
\zd_2 q_2) \in \sT_\oq \oQ$, and $\zd_1\oq = (0, \zd_1q_2) \in \sV_\oq \oQ$.  A mapping $\zq \,\colon \R^2 \rightarrow \oQ $ can be choosen of the
form
        $$ \zq(s_1, s_2) = (\zq_1(s_1, s_2), \zq_2(s_1)),
                                                                                                           \eqno(151)$$
    where $\zq_1$ represents the pair $\zd_1 q_1, \zd_2 q_1 \in \sT_{q_1} Q$, and $\zq_2$ represents the vector $\zd_2
q_2 \in \sT_{q_2}Q$. We have from \RF(149) and \RF(150)
        $$ \sH(\oU,\zh,\oq)(\zd_1\oq,\zd_2\oq) =  \rD^{(1,1)}(\oU
\comp \zq)(0,0) = \langle g(\zd_1 q_2),\zd_2 q_2 - \zd_2 q_1 \rangle .
                                                                                                           \eqno(152)$$
        }
        \Example{
    Here, we consider the generating family of Example~6. At
        $$\oq = (q_1,q_2) \in \sC\sr(\oU,\zh) = \{(q_1,q_2)\in Q\times Q ;\;  \|q_1 - q_2\| = 0\},
                                                                                                           \eqno(153)$$
    we have
        $$ \sH(\oU,\zh,\oq)(\zd_1\oq,\zd_2\oq) = k_2 {{1}\over{a^2}} \langle g(q_2 - q_1), \zd_2 q_2 - \zd_2 q_1 \rangle
\langle gq(q_2 - q_1),\zd_1 q_2 \rangle .
                                                                                                           \eqno(154)$$
        }

        \sect{Regular families of generating functions.}

    Let

    \vskip1mm
        $$\vcenter{
    \begindc{0}[1]
    \obj(000,055)[01]{$\oQ$}
    \obj(000,000)[00]{$Q$}
    \mor{01}{00}[10,10]{$\zh$}[-1,0]
    \enddc}
        $$
    \vskip2mm\noindent
    be a differential fibration, let $\oq$ be a point in $\oQ$ and let $\of$ be an element of $\sV_\oq^\polar\oQ$.  We
choose a function $F \,\colon Q \rightarrow \R$ such that $\rd_{\zh(\oq)}F =\t\zh(\of)$ and use the function $\oF = F\comp \zh$ to define a spliting
$\sT_\of\sT^\* \oQ = H_\of + V_\of$ at $\rd\oF(\oq) = \of$.  Note that $\rd \oF(\oQ) \subset \sV^\polar\oQ$.  Hence,

        $$i_h(\sT_\oq\oQ) = H_\of = \sT\rd\oF(\sT_\oq\oQ) \subset \sT_\of\sV^\polar\oQ.
                                                                                                           \eqno(155)$$
    The equality

        $$i_v(\sV^\polar_\oq \oQ) = V_\of \cap \sT_\of\sV^\polar\oQ
                                                                                                           \eqno(156)$$
    is a consequence of general properties of the injection $i_v$.  The two equalities \RF(155) and \RF(156) result in
        $$\zF(\sT_\oq\oQ \oplus \sV^\polar_\oq \oQ) = \sT_\of\sV^\polar\oQ.
                                                                                                           \eqno(157)$$

    The space $\sV_\oq \oQ \oplus \{0\}$ is the symplectic polar of $\sT_\oq \oQ\oplus \sV^\polar_\oq \oQ$ in the
symplectic space $\sT_\oq \oQ\oplus \sT^\*_\oq \oQ$.  Hence,
        $$(\sT_f\sV^\polar \oQ)\srP = i_h(\sV_\oq \oQ) = \sT\rd\oF(\sV_\oq \oQ).
                                                                                                           \eqno(158)$$
    This convenient expression for the symplectic polar is obviously independent of the choice of the function $F$.

    Let $\oq \in \sC\sr(\oU,\zh)$ be a critical point of a family $(\oU,\zh)$, let $\oS$ be the Lagrangian submanifold
$\rd\oU(\oQ)$ and let $\of = \rd \oU(\oq) \in \oS$.  Let $F$ be one of the functions on $Q$ used in Section 8 to define the Hessian $\sH(\oU,\zh, \oq)$
at $\oq$.  The function $\oF = F \comp \zh$ is used to construct an isomorphism
        $$\zC \,\colon \sT_\of\sT^\*Q \rightarrow \sT_\oq\oQ \oplus \sT_\oq^\*\oQ.
                                                                                                           \eqno(159)$$

    The space $\sT_f\oS \subset \sT_\of^\*\oQ$ is Lagrangian subspace.  Its image
        $$L_\of = \zC(\sT_\of\oS) \subset \sT_\oq\oQ \oplus \sT_\oq^\*\oQ
                                                                                                           \eqno(160)$$
    is the graph of a symmetric linear mapping
        $$\zl_\of \,\colon \sT_\oq\oQ \rightarrow \sT_\oq^\*\oQ.
                                                                                                           \eqno(161)$$
    We have
        $$\sH(\oU - \oF, \oq)(\zd_1\oq, \zd_2\oq) = \langle \zl_\of(\zd_1 \oq), \zd_2 \oq \rangle
                                                                                                           \eqno(162)$$
    We introduce a rather obvious definition
        $$\ker\sH(\oU-\oF,\oq) = \ker \zl_\of
                                                                                                           \eqno(163)$$
    and a less obvious definition
        $$\ker \sH(\oU,\zh, \oq) = \ker \zl_\of \cap \sV_\oq \oQ.
                                                                                                           \eqno(164)$$
    We have then
        $$ \ker \sH(\oU,\zh,\oq) = \{\zd \oq \in\sV_\oq \oQ ;\; i_h(\zd \oq) \in \sT_f\oS \}
                                                                                                           \eqno(165)$$
        and
        $$\eqalign{
    i_h(\ker \sH(\oU,\zh,\oq)) &= i_h(\sV_\oq) \cap \sT_f \oS \cr
    &= \ker \sT_f \t\zh \cap \sT_f \oS \cr
    }
                                                                                                           \eqno(166)$$
    Consequently,
        $$\dim (\ker \sT_f \t\zh \cap \sT_f \oS) = \dim (\ker \sH(\oU,\zh,\oq)) = \dim (\sV_\oq \oQ) - \tx{rank}\,
\sH(\oU,\zh, \oq)
                                                                                                           \eqno(167)$$

        \Definition{{\rm
    A family $(\oU,\zh)$ is called a {\it Morse family} if the rank of $\sH(\oU,\zh,\oq)$ is maximal at each $\oq \in~
\sC\sr(\oU,\zh)$.  The family $(\oU,\zh)$ is said to be {\it regular} if the critical set $\sC\sr(\oU,\zh)$ is a submanifold of $\oQ$ and the rank of
$\sH(\oU,\zh,\oq)$ at each $\oq \in \sC\sr(\oU,\zh)$ is equal to the codimension of $\sC\sr(\oU,\zh)$.
    }}

    We will show that a regular family generates a Lagrangian submanifold of $\sT^\* Q$ and that a Morse family is
regular.

        \Theorem{
    If $(\oU,\zh)$ is a regular family, then the image of $\zk(\oU,\zh)$ is an immersed Lagrangian submanifold of
$\sT^\* Q$.
    }
        \Proof{
    Let $\oq \in \sC\sr(\oU,\zh)$ and $\of =\rd \oU(\oq)$.  The rank of $\zk(\oU,\zh)$ at $\oq$ is equal to
        $$\dim(\sT_\oq \sC\sr(\oU,\zh)) - \dim(\ker(\sT_\oq\zk(\oU,\zh))).
                                                                                                           \eqno(168)$$
    We have
        $$\eqalign{
    \ker (\sT_\oq \zk(\oU,\zh)) &= \ker(\sT_\of \t{\zh}) \cap \sT_\of \rd \oU (\sC\sr(\oU,\zh)) \cr
    &= \ker(\sT_\of \t{\zh}) \cap \sT_\of(\oS\cap V^\polar \oQ)\cr
    &\subset \ker(\sT_\of \t{\zh}) \cap (\sT_\of \oS)\cap \sT_\of(V^\polar \oQ) \cr
    &= \ker(\sT_\of \t{\zh}) \cap \sT_\of \oS.}
                                                                                                           \eqno(169)$$
    It follows from \RF(169) and from \RF(167) that
        $$\dim(\ker(\sT_\oq\zk(\oU,\zh))) \leqs \dim (\ker(\sT_\of \o{\zh}) \cap \sT_\of \oS) = \dim \sV_\oq\oQ - {\rm
rank}\, \sH(F,\zh,\oq).
                                                                                                           \eqno(170)$$
    Since the family $(\oU,\zh)$ is regular, ${\rm rank}\, \sH(\oU,\zh,\oq) = \dim \sV_\oq\oQ + \dim Q - \dim
\sC\sr(\oU,\zh)$ and, consequently,
        $$ \dim(\ker(\sT_\oq\zk(\oU,\zh))) \leqs \dim \sC\sr(\oU,\zh) - \dim Q.
                                                                                                           \eqno(171)$$
    It follows that
        $$ \dim(\im(\sT_\oq\zk(\oU,\zh))) = \dim \sC\sr(\oU,\zh) - \dim(\ker(\sT_\oq\zk)) \geqs \dim Q.
                                                                                                           \eqno(172)$$
    \dacapo
    On the other hand, $\sT_\oq\zk(\oU,\zh)$ is the composition of $\sT_\oq\rd \oU$, restricted to
$\sT_\oq\sC\sr(\oU,\zh)$, and the strict symplectic reduction $\sT_\of \t{\zh}$, which is the essential part of the symplectic reduction relation
        $$\sT_\of\sPh\zh \,\colon \sT_\of\sT^\* \oQ\rightarrow \sT_{\o{\zh}(f)}\sT^\* Q.
                                                                                                           \eqno(173)$$
    The image $\sT_y\rd \oU(\sT_\oq \sC\sr(\oU,\zh))$ is an isotropic subspace of $\sT_\of\sT^\* \oQ$ and, consequently,
$\im(\sT_\oq\zk(\oU,\zh))$ is an isotropic subspace of $\sT_{\t\zh(f)}\sT^\* Q$.  This implies the inequality
        $$\dim(\im(\sT_\oq\zk) \leqs \dim Q,
                                                                                                           \eqno(174)$$
    and, consequently,
        $$\dim(\im(\sT_\oq \zk)) = \dim(Q).
                                                                                                           \eqno(175)$$
    It follows from the constant rank theorem that $S = \zk(\oU,\zh)(\sC\sr(\oU,\zr))$ is an immersed submanifold of
$\sT^\* Q$ and $\dim(S) = \dim(Q)$.  Since $S$ is isotropic it is Lagrangian.
    }

        \Proposition{
    A Morse family is regular.
        }

        \Proof{
    We have to show that the critical set of a Morse family $(\oU,\zh)$ is a submanifold of dimension $\dim Q$.
    Let $\oq$ be a critical point of the family, $\of= \rd \oU(\oq)$ and
        $$\zC \,\colon \sT_\of\sT^\*\oQ \rightarrow \sT_\oq \oQ\oplus \sT^\*_\oq \oQ
                                                                                                           \eqno(176)$$
    the isomorphism constructed with a function on $Q$ as in Section 8.  The image $L_\of = \zC(\sT_\of\oS)$ of
$\sT_\of\oS$ is the graph of a symmetric mapping $\zl_\of \,\colon \sT_\oq \oQ \rightarrow \sT^\*_\oq \oQ$.  The rank of the Hessian of $(\oU, \zh)$ at
$\oq$ is the rank of $\zl_\of$ restricted to $\sV_\oq \oQ$.  Let $\zl_{(\of,v)} \,\colon \sV_\oq \oQ \rightarrow \sT^\*_\oq \oQ$ be this restriction.
The dual mapping $\zl_{(\of,v)}^\* \,\colon \sT_\oq \oQ \rightarrow \sV^\*_\oq \oQ$ is of the same rank.  Since $\zl_\of$ is symmetric,
$\zl_{(\of,v)}^\* = \zr_\oq \comp \zl_\of $, where $\zr_\oq$ is the restriction of the canonical projection
        $$\zr \,\colon \sT^\*\oQ \rightarrow \sV^\* \oQ
                                                                                                           \eqno(177)$$
    to $\sT^\*_\oq \oQ$.  The injections $i_h, i_v$ induce injections $i_{(h,\zr)} \,\colon \sT_\oq \oQ \rightarrow
\sT_{\zr(f)}\sV^\* \oQ$ and $i_{(v,\zr)} \,\colon \sV_\oq^\* \oQ \rightarrow \sT_{\zr(f)}\sV^\* \oQ$ and an isomorphism
        $$\zC_\zr \,\colon\; \sT_{\zr(f)}\sV^\* \oQ \rightarrow \sT_\oq \oQ \oplus \sV^\*_\oq \oQ.
                                                                                                           \eqno(178)$$
    With this isomorphism, the mapping $\sT_\oq(\zr\comp \rd \oU )$ is represented by $\zr_\oq \comp \zl_\of=
\zl_{(\of,v)}^\*$.  The mapping $\zr \comp \rd \oF$ is the zero section of $\sV^\*\oQ$.  It follows that the image of $i_{(h,\zr)}$ is tangent to the
zero section.
    We choose a local trivialization
        $$ \zz \,\colon \sV^\*_O \oQ \rightarrow \sV^\*_\oq \oQ
                                                                                                           \eqno(179)$$
    of $\sV^\*\oQ$ in a neighbourhood $O$ of $\oq$.
    We have $\sC\sr(\oU, \zh)\cap O = (\zz\comp \zr \comp \rd \oU)^{-1}(0)$ and $\sT_\oq(\zz \comp \zr \comp \rd \oU)
\colon \sT_\oq \oQ \rightarrow \sV^\*_\oq \oQ$ coincides with $\zl_{(\of,v)}^\*$.  The rank of the Hessian of the family $(\oU,\zh)$ at $\oq$ is the
rank of $\zl_{(\of,v)}^\*$ and consequently, the rank of $\sT_\oq(\zz\comp \zr \comp\rd \oU)$.  It is maximal, hence equal $\dim \sV_\oq \oQ = \dim
\sV^\*_\oq \oQ$.  It follows that $\sT_\oq(\zz \comp\rd \oU)$ is surjective and, by the implicit function theorem, $\sC\sr(\oU,\zh)$ is a submanifold
of dimension $\dim\oQ - \dim \sV^\*_\oq \oQ = \dim Q$.
        }

        \Proposition{
    The family $(\oU,\zh)$ is regular if and only if $\oS = \rd \oU (\oQ)$ and $\sV^\polar \oQ$ have clean intersection.
    }
        \Proof{
    Let $\of\in \oS\cap \sV^\polar \oQ$ and $\zp_\oQ(\of) =\oq$.  As in the preceding proposition, we shall use the
canonical projection \RF(177) and the isomorphism \RF(178).
    We have
        $$\zC_\zr(\sT\zr (\sT_\of\sV^\polar \oQ))= \sT_\oq \oQ \oplus {0}
                                                                                                           \eqno(180)$$
    and $\zC_\zr(\sT_\oq (\zr \comp \rd \oU))$ is the graph of $\zr_\oq \comp \zl_\of =\zl^\*_{(\of,v)} \,\colon \sT_\oq
\oQ \rightarrow \sV^\*_\oq$.  The rank of $\zl_{(\of,v)}$ is equal to the rank of the Hessian of the family $(\oQ,\zh)$ at $\oq$.  It follows that
        $$\sT \zr(\sT_\of\sV^\polar \oQ + \sT_\of \oS) = \sT_\oq \oQ \oplus \im (\zl_{(\of,v)}),
                                                                                                           \eqno(181)$$
    and the dimension of these spaces is $\dim \oQ + \dim \im(\zl_\of)= \dim \oQ + {\rm rank}\, \sH(\oU,\zh,\oq)$.
Since the kernel of $\sT_\of\zr$ is contained in $\sT_\of\sV^\polar \oQ$, we have
        $$\eqalign{
    \dim(\sT_\of\sV^\polar \oQ + \sT_\of \oS) &= \dim (\sT \zr(\sT_\of\sV^\polar \oQ + \sT_\of \oS)) + \dim \ker
(\sT_\of \zr) \cr
    &= \dim \oQ + {\rm rank}\, \sH(\oU, \zh,\oq) + \dim Q \cr
    }
                                                                                                           \eqno(182)$$
    It follows that
        $$\eqalign{
    \dim(\sT_\of\sV^\polar \oQ \cap \sT_\of \oS) &= \dim(\sT_\of\sV^\polar \oQ)+ \dim(\sT_\of \oS) -
\dim(\sT_\of\sV^\polar \oQ + \sT_\of \oS) \cr
    &= \dim(\oQ) + \dim(Q) + \dim(\oQ) - \dim(\oQ) - \dim(Q) -{\rm rank}\, \sH(\oU, \zh,\oq) \cr
    &= \dim(\oQ) -{\rm rank}\, \sH(\oU, \zh,\oq) \cr.
    }
                                                                                                           \eqno(183)$$
    We conclude that
    $ \sT_\of \sC\sr(\oU, \zh) = \sT_\of \sV^\polar \oQ \cap \sT_\of \oS$ if and only if $\dim (\sC\sr(\oU,\zh)) =
\dim(\oQ) - \;{\rm rank}\,\sH(\oU, \zh,\oq).$
    }

        \Corollary{
    $(\oU, \zh)$ is a Morse family if and only if $\sV^\polar R$ and $\oS$ have transversal intersection.
    }
        \Proof{
    We have from \RF(182) that $\dim(T_\of \sV^\polar \oQ + \sT_\of\oS) = \dim \sT^\* \oQ$ if and only if
        $$\dim(Q) + \;{\rm rank}\, ( \sH(\oU, \zh,\oq)) = \dim(\oQ),
                                                                                                           \eqno(184)$$
    i.e., if and only if $ \sH(\oU, \zh,\oq)$ is of maximal rank.
    }


        \Example{
    Let $(\oU,\zh)$ be the generating family of Example~5. The Hessian of this family
        $$  \sH(\oU,\zh,\oq)(\zd_1\oq,\zd_2\oq) =  \rD^{(1,1)}(\oU
\comp \zq)(0,0) = \langle g(\zd_1 q_2),\zd_2 q_2 - \zd_2 q_1 \rangle
                                                                                                           \eqno(185)$$
    is of maximal rank.  The family is a Morse family.
        }
        \Example{
    For the family $(\oU,\zh)$ of Example~6, the critical set
        $$ \sC\sr(\oU,\zh) = \left\{(q_1,q_2) \in Q \times Q ;\; \|q_2 - q_1\| = a \right\}
                                                                                                           \eqno(186)$$
    is a submanifold of codimension 1. The Hessian
        $$ \sH(\oU,\zh,\oq)(\zd_1\oq,\zd_2\oq) = k_2 {{1}\over{a^2}} \langle g(q_2 - q_1), \zd_2 q_2 - \zd_2 q_1 \rangle
\langle gq(q_2 - q_1),\zd_1 q_2 \rangle
                                                                                                           \eqno(187)$$
    is of constant rank 1. The family is regular.
        }

        \Example{{\rm
    Let $\oQ= \R^2$, $Q= \R$ and $\zh\colon \R^2\rightarrow \R \,\colon (x,\zl) \mapsto x$.  For $\oU(x,\zl) = \zl x^2$
we have $\sC\sr(\oU,\zh)= \{(x,\zl)\colon x=0\}$ and the Hessian is the trivial zero form.  In this case the intersection of $\oS$ and $\sV^\polar\oQ$
is not clean, but the Hessian is of constant rank.  The generated set is an isotropic submanifold, but not Lagrangian.
    }}

    \sect{References.}

    \item{[1]} P. Libermann and Ch.-M.~Marle, {\it Symplectic Geometry and Analytical Mechanics}, Reidel, Dordrecht 1987.
    \item{[2]} G. Pidello and W. M. Tulczyjew, {\it Derivations of differential forms on jet bundles}, Ann.Mat.Pura
Appl. {\bf 147} (1987), 249-–265.
   \item{[3]} W. M. Tulczyjew, {\it Hamiltonian Systems, Lagrangian Systems and the Legendre Transformation}, Symposia Mathematica, {\bf 16} (1974),
       247--258.

    \item{[4]}  W. M. Tulczyjew and P. \ Urba\'nski, {\it Liouville structures}, Universitatis Iagellonicae Acta Mathematica,
    {\bf 47} (2009), 187--226,   arXiv:0806.1333.

    \item{[5]} A. Weinstein, {\it Lectures on symplectic manifolds}, CBMS regional conference series in Mathematics {\bf 29}, American Mathematical
        Society, Providence, 1977.
    \end